\UseRawInputEncoding

\documentclass[prb,twocolumn,superscriptaddress,amsmath,amssymb,longbibliography]{revtex4-2}

\usepackage{graphicx}
\usepackage{latexsym}
\usepackage{amsmath}
\usepackage{amssymb}
\usepackage{amsfonts}
\usepackage{color}
\usepackage{bm}
\usepackage{verbatim}
\usepackage{pagecolor,lipsum}
\usepackage{float}
\usepackage{hyperref}
\usepackage[normalem]{ulem}
\hypersetup{colorlinks=true,urlcolor=blue,linkcolor=blue,citecolor=blue}
\usepackage{comment}
\usepackage{siunitx}

\begin{document}


\title{Beyond the effective length: How to analyze magnetic interference patterns of thin film planar Josephson junctions with finite lateral dimensions}

\date{\today}

\author{R. Fermin}
\author{B. de Wit}
\author{J. Aarts}
\affiliation{Huygens-Kamerlingh Onnes Laboratory, Leiden University, P.O. Box 9504, 2300 RA Leiden, The Netherlands.}



\begin{abstract}
The magnetic field dependent critical current $I_{\text{c}}(B)$ of a Josephson junction is determined by the screening currents in its electrodes. In macroscopic junctions, a local vector potential drives the currents, however, in thin film planar junctions, with electrodes of finite size and various shapes, they are governed by non-local electrodynamics. This complicates the extraction of parameters such as the geometry of the effective junction area, the effective junction length and, the critical current density distribution from the $I_{\text{c}}(B)$ interference patterns. Here we provide a method to tackle this problem by simulating the phase differences that drive the shielding currents and use those to find $I_{\text{c}}(B)$. To this end, we extend the technique proposed by John Clem [Phys. Rev. B, \textbf{81}, 144515 (2010)] to find $I_{\text{c}}(B)$ for Josephson junctions separating a superconducting strip of length $L$ and width $W$ with rectangular, ellipsoid and rhomboid geometries. We find the periodicity of the interference pattern ($\Delta B$) to have geometry independent limits for $L \gg W$ and $L \ll W$. By fabricating elliptically shaped S$-$N$-$S junctions with various aspect ratios, we experimentally verify the $L/W$ dependence of $\Delta B$. Finally, we incorporate these results to correctly extract the distribution of critical currents in the junction by the Fourier analysis of $I_{\text{c}}(B)$, which makes these results essential for the correct analysis of topological channels in thin film planar Josephson junctions.
\end{abstract}

 \maketitle

\section{Introduction}


Planar Josephson junctions are ubiquitous in modern solid state physics research, with examples ranging from topological junctions\cite{Hart2014,Pribiag2015,Fornieri2019}, high $T_{\text{c}}$ (grain boundary) junctions\cite{Mayer1993,Cybart2015a}, gated-junctions that control supercurrent flow\cite{Ying2020,Elfeky2021}, graphene-based junctions\cite{Allen2016,Fortin2022}, magnetic field sensors\cite{Golod2019,LeFebvre2022,Hovhannisyan2022} and, junctions with a ferromagnetic weak link\cite{Lahabi2017a,Co_disk_paper,Jeon2021}. A major tool in analysing these junctions experimentally is the magnetic interference pattern observed in the critical current ($I_{\text{c}}(B)$), the shape and periodicity of which can reveal, using Fourier transform, information about the underlying distribution of critical current in the weak link\cite{Dynes1971}. Often this Fourier analysis is carried out in terms of an effective junction length, given, for macroscopic junctions, by $2\lambda + d$, where $\lambda$ is the London penetration depth and $d$ the thickness of the weak link. This effective length originates from the Meissner effect. However, when the junction is formed between two superconducting thin films, with a thickness below $\lambda$, the shielding currents running along the junction, responsible for the shape and periodicity of the magnetic interference of the critical current $I_{\text{c}}(B)$, are no longer determined by the Meissner effect in its macroscopic form (i.e., by the local vector potential). Rather they are determined by non-local electrodynamic effects\cite{Pearl1964,Ivanchenko1990,Abdumalikov2009,Boris2013}. 

In numerous theoretical and experimental studies, it was found that in thin film planar junctions, $I_{\text{c}}(B)$ becomes completely independent of $\lambda$ and is solely determined by the geometry of the sample\cite{Kogan2001,Moshe2008,Clem2010,Boris2013,Rodan-Legrain2021}. Moreover, John Clem provided a method to calculate $I_{\text{c}}(B)$ for planar junctions that are also restricted in their lateral size (i.e., a Josephson junction separating a rectangular superconducting strip of width $W$ and length $L$ in two halves)\cite{Clem2010}. As experimental studies often deal with finite-size geometries, his theory is highly topical at the moment.

This paper bridges the gap between predicting the $I_{\text{c}}(B)$ of thin film planar junctions featuring finite lateral geometry, and the correct analysis of the experimental interference patterns used to extract the current density distribution. First we review the technique proposed by Clem and extend on his work by covering two more geometries: the ellipse and the rhomboid. We calculate $I_{\text{c}}(B)$ for these geometries, extract the periodicity of the interference pattern ($\Delta B$) for different ratios of $L/W$, and find $\Delta B$ to have two geometry independent limits for $L \gg W$ and $L \ll W$. By fabricating elliptically shaped S$-$N$-$S junctions with different ratios of $L/W$, we experimentally verify the geometry dependence of $\Delta B$. Finally, we adapt the well-known Fourier relation between $I_{\text{c}}(B)$ and the critical current density distribution for use on laterally finite thin film planar junctions. We find that altering the Fourier transform is crucial for predicting the location of possible current channels in thin film planar junctions.

\section{Review of the Clem model}

\begin{figure}[t!]
 \centerline{$
 \begin{array}{c}
  \includegraphics[width=1\linewidth]{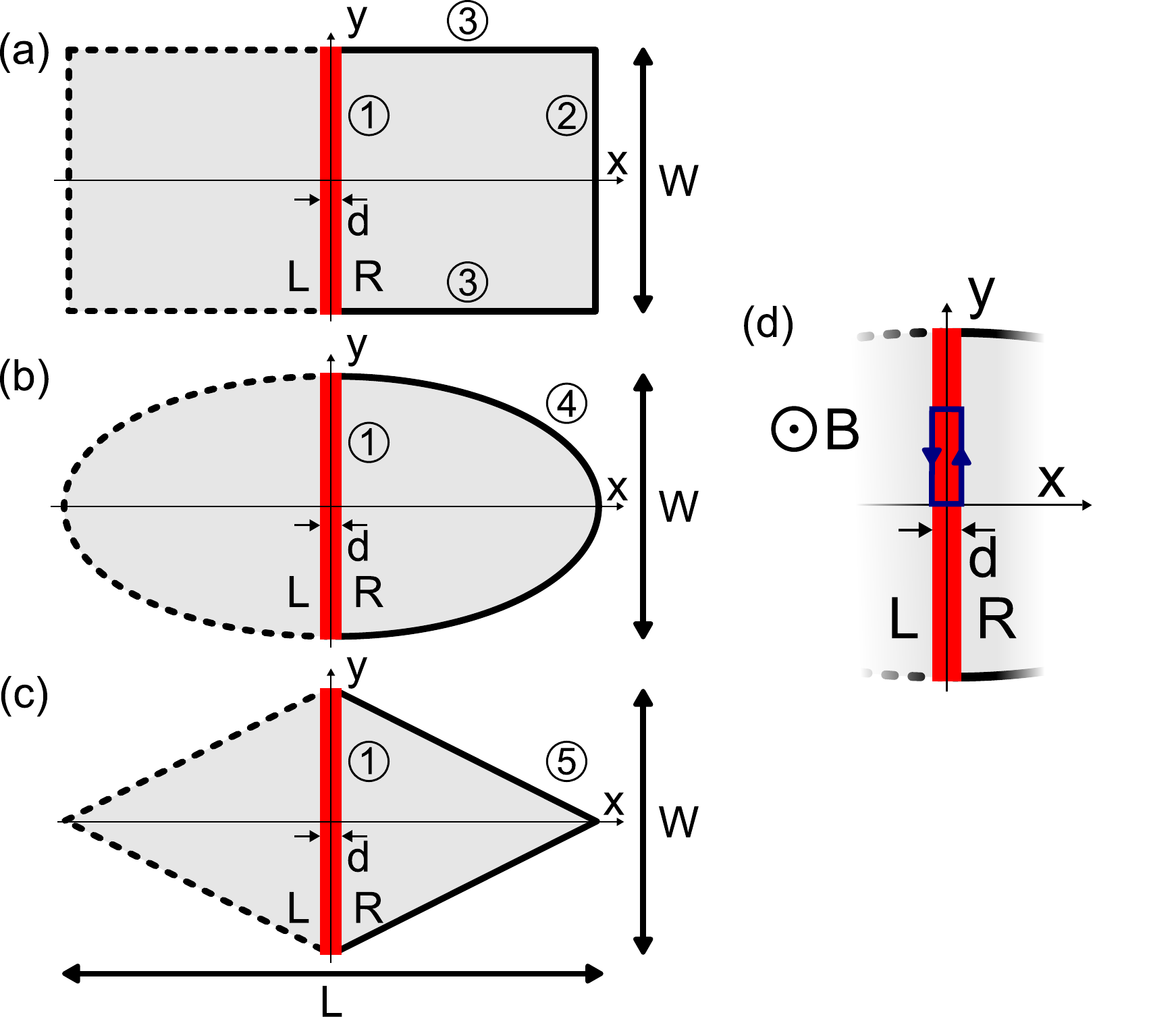}
 \end{array}$}
 \caption{Schematics of the three geometries used for calculating $I_{\text{c}}(B)$, being (a) the rectangle, (b) the ellipse and (c) the rhombus. The schematics resemble superconducting thin films of width $W$ and length $L$, which are separated by a normal metal junction of width $d$ (colored red). By numbers we indicate different sections of the right electrode edge. The boundary conditions of the calculations for these are summarized in Table \ref{table1}. In (d) we show a zoom of the junction area under the magnetic induction $\pmb{B} = B\hat{z}$. The dark blue path is used as loop integral to determine $I_{\text{c}}(B)$.} \label{H3_sim_geom}
\end{figure}

We consider a normal metal Josephson junction (dimensions $W_{\text{JJ}}$ and $d$) that divides a symmetric superconducting thin film, having dimensions $L$ and $W$, into two halves. Figure \ref{H3_sim_geom} shows a schematic of three of such films, having different geometries. The junction, colored red in Figure \ref{H3_sim_geom}, is running along the y-direction from $-W/2$ to $W/2$ (i.e., $W_{\text{JJ}} = W$). Since we examine the thin film limit, the screening current density is assumed uniform along the thickness of the film, which effectively reduces the problem to a 2D one. We specifically consider the junction to be in the short junction limit, as the model by Clem treats an infinitesimally thin insulating tunnel junction. Furthermore, it is assumed that the electrode the electrode dimensions are smaller than the Pearl length, given by:

\begin{align} \label{Eq:H3_JPL1}
\Lambda = \frac{2\lambda^2}{t_{\text{film}}}
\end{align}

\noindent Where $t_{\text{film}}$ the thickness of the superconducting films. This implies that the self fields originating from the screening currents are far smaller than the applied external field. Additionally we assume that the junction is in the narrow limit, meaning that the junction is less wide than the Josephson penetration length, which for planar junctions in the thin film limit is the given by\cite{Kogan2001,Clem2010,Boris2013}:

\begin{align} \label{Eq:H3_JPL2}
l = \frac{\Phi_0 t_{\text{junc}} W}{4\pi\mu_0\lambda^2I_{\text{c}}(0)}
\end{align}

\noindent Here $t_{\text{junc}}$ is the thickness of the junction (not necessarily equal to the thickness of the film), $I_{\text{c}}(0)$ its critical current at zero magnetic field, $\mu_0$ is the vacuum permeability, and $\Phi_0$ is the magnetic flux quantum.

In order to calculate $I_{\text{c}}(B)$, we assume a sinusoidal current-phase relation $J_{x}= J_{\text{c}} \sin \varphi (y)$, where $\varphi(y)$ is the gauge-invariant phase difference over the junction, which depends on the location along the junction. It can be evaluated within the framework of Ginzburg-Landau theory by considering the second Ginzburg-Landau equation, which is given as:

\begin{align} \label{Eq:4}
\pmb{J} = -\frac{\Phi_0}{2\pi\mu_0\lambda^2}\left(\frac{2\pi}{\Phi_0}\pmb{A} + \nabla \gamma \right) = \frac{\Phi_0}{2\pi\mu_0\lambda^2} \theta
\end{align}

\noindent Here $\pmb{A}$ is the vector potential corresponding to the applied magnetic field ($\pmb{B} = \pmb{\nabla} \times \pmb{A}$), and $\gamma$ is the gauge covariant phase of the wavefunction describing the superconducting order parameter (given by $\Psi = \Psi_0 e^{i\gamma}$\footnote{Here we assume a weak Josephson current, such that the magnitude of the superconducting order parameter is not suppressed, and is given by the equilibrium value.}). Finally, $\theta$ is the gauge-invariant phase gradient (required by the fact that \pmb{J} is a gauge-invariant property). $\varphi (y)$ is then given by integrating $\theta$ across the junction:

\begin{align} \label{Eq:15}
\varphi(y) = \gamma(-\frac{d}{2},y)-\gamma(\frac{d}{2},y)-\frac{2\pi}{\Phi_0} \int_{-d/2}^{d/2} A_x(x,y) \: \mathrm{d}x
\end{align}

In Figure \ref{H3_sim_geom}d, we sketch a zoom of a junction, where we specify an integration contour under a magnetic induction of $\pmb{B} = B\hat{z}$. By integrating $\pmb{\nabla}\gamma$ along this contour and realizing that $\int_{\text{C}} \pmb{\nabla} \gamma \: \mathrm{d}\pmb{l}  = 2\pi n$, where $n$ is an integer and $\sin{(\varphi+2\pi n)} = \sin{(\varphi)}$, we find:

\begin{align} \label{Eq:19}
\varphi(y) = \varphi(0) + \frac{2\pi}{\Phi_0} \left(y d B + 2\mu_0\lambda^2 \int_0^y  J_{y}(\frac{d}{2},y') \: \mathrm{d}y' \right)
\end{align}

\begin{center}
\begin{table}
\bgroup
\def\arraystretch{2}
\begin{tabular}{|c|c|}
\hline
Boundary & $(\pmb{\nabla}\gamma)\cdot\pmb{\hat{n}}_\Omega$  \\[2pt]
\hline\hline
1 & $\frac{2\pi B}{\Phi_0}y$  \\[2pt]
\hline
2 & $-\frac{2\pi B}{\Phi_0}y$  \\[2pt]
\hline
3 & 0\\
\hline
4 & $\frac{2\pi B}{\Phi_0}\frac{W x y}{L\sqrt{(\frac{Wx}{L})^2+(\frac{Ly}{W})^2}}$  \\[8pt]
\hline
5 & $-\frac{2\pi B}{\Phi_0}\frac{W y}{\sqrt{W^2+L^2}}$\\[2pt]
\hline
\end{tabular}
\egroup
\caption{The Neumann boundary conditions for each electrode boundary, listed by the numbering used in Figure \ref{H3_sim_geom}.}\label{table1}
\end{table}
\end{center}

\begin{figure*}[t!]
 \centerline{$
 \begin{array}{c}
  \includegraphics[width=0.8\linewidth]{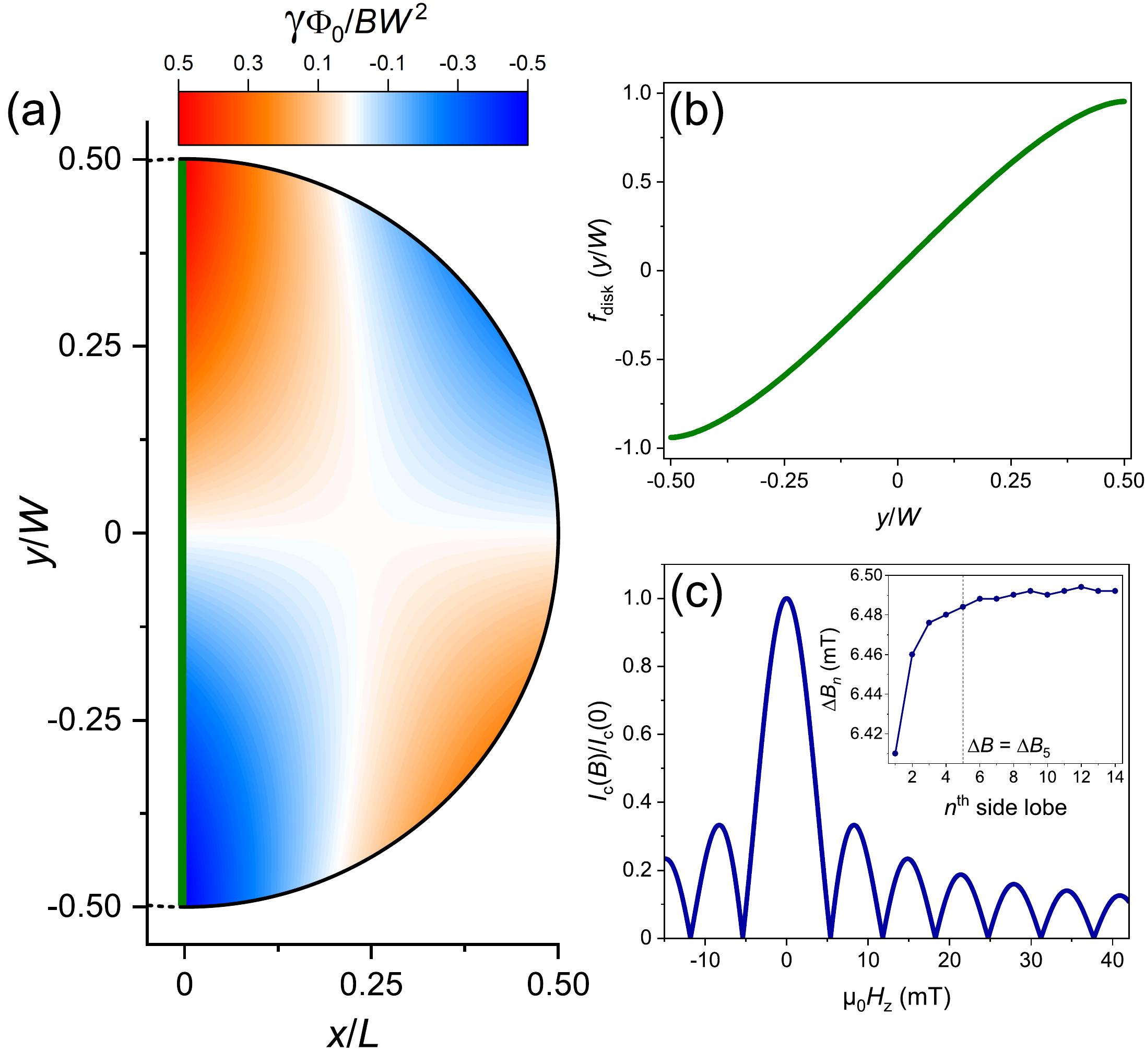}
 \end{array}$}
 \caption{(a) Gauge-covariant phase simulated in the right electrode for a disk-shaped planar Josephson junction, normalized to the applied magnetic field and width of the junction $\gamma \Phi_0 / BW^2$. The junction is shown as a green line. This result allows for extracting the gauge-covariant phase along the junction. It follows the scaling of Eq. \ref{Eq:H3_7}, and it is determined by a dimensionless function, which is plotted in (b). (c) Shows the interference pattern calculated using the result in (a) by numerically evaluating Equation \ref{Eq:H3_6} for different values of $B$. The typical interference pattern looks like a Fraunhofer pattern at first sight. However, the peak height decreases less strongly than $1/B$, and the width of the side lobes is larger than half of the middle lobe, which is 10.76 mT wide. Furthermore, the width of the $n$th side lobe increases and reaches an asymptotic value for large values of $n$, which is evident from the inset of (c), where we plot the width of the $n$th side lobe. The width of the fifth side lobe is used for comparisons between simulations and experiments.} \label{H3_disk_results}
\end{figure*}

\noindent Here we have used Stokes theorem to evaluate the flux entering the contour and used the fact that the electrodes are mirror symmetric ($J_{y}(\frac{d}{2},y)$ = -$J_{y}(-\frac{d}{2},y)$).
For macroscopic junctions $J_{y,R}(\frac{d}{2},y') = \frac{B\mu_0}{\lambda_{\text{L}}}$ resulting from the the Meissner effect, leading to $\varphi(y) = \varphi(0) + \frac{2\pi(2\lambda+d)B}{\Phi_0}y$, where we recognize the effective junction length. Since the junctions considered here are in the thin film limit, we take a different approach in evaluating $J_{y}(\frac{d}{2},y')$. First note that the supercurrent is conserved and therefore $\pmb{\nabla}\cdot\pmb{J} = 0$. By choosing the convenient gauge $\pmb{A} = -yB\hat{x}$, we find $\nabla \times \pmb{A} = B\hat{z}$ and $\nabla \cdot \pmb{A} = 0$. Therefore, the divergence of the second Ginzburg-Landau equation (Eq. \ref{Eq:4}) reduces to:

\begin{align} \label{Eq:H3_1}
\nabla^2 \gamma = 0
\end{align}\textbf{}

\noindent Therefore, we mapped the second Ginzburg-Landau equation onto the Laplace equation. With sufficient boundary conditions, it can be solved for a unique solution, which allows us to calculate $J_y(\frac{d}{2},y)$. The boundary conditions arise from the prerequisite that no supercurrent can exit the sample at its outer boundaries. Furthermore, we assume a weak Josephson coupling, meaning that the shielding currents in the electrodes are far larger than the Josephson currents between the electrodes, which we approximate as $J_x(\frac{d}{2},y)=0$. Therefore, we can write:

\begin{align} \label{Eq:H3_2}
\pmb{J}\cdot\pmb{\hat{n}}_{\text{R}} = 0 
\end{align}

\noindent Where $\pmb{\hat{n}}_{\text{R}}$ is the unit vector, normal to the outer edges of the right electrode. Combined with the second Ginzburg-Landau equation, this leads to a set of Neumann boundary conditions:

\begin{align} \label{Eq:H3_3}
(\pmb{\nabla}\gamma)\cdot\pmb{\hat{n}}_{\text{R}} = -\frac{2\pi}{\Phi_0}\pmb{A}\cdot\pmb{\hat{n}}_{\text{R}}
\end{align}

\noindent Which is sufficient to solve for $\gamma(x,y)$. Next, Eq. \ref{Eq:19} allows us to find the gauge-invariant phase difference over the junction $\varphi (y)$. Note that we have conveniently chosen $A_{\text{y}} = 0$. We then find:

\begin{align} \label{Eq:H3_4}
2\mu_0\lambda^2 \int_0^y  J_{y}\left(\frac{d}{2},y'\right) \: \mathrm{d}y' = 2 \gamma\left(\frac{d}{2},y\right)
\end{align}

\noindent Therefore, $\varphi (y)$ is given by the simple expression:

\begin{align} \label{Eq:H3_5}
\varphi (y) = \varphi(0) + \frac{2\pi dB}{\Phi_0}y + 2 \gamma\left(\frac{d}{2},y\right)
\end{align}

\noindent Next, the current across the junction is given by $\int \pmb{J} \: \mathrm{d}\pmb{S}$, yielding:

\begin{align} \label{Eq:21}
I(B) =  \int_{-W/2}^{W/2} t_{\text{junc}} J_{\text{c}} \sin{\left(\varphi(0) + \frac{2\pi dB}{\Phi_0}y + 2 \gamma\left(\frac{d}{2},y\right)\right)} \: \text{d}y 
\end{align}

\noindent We assume that the critical current density at zero field is distributed uniformly over the junction, yielding $J_{\text{c}} = \frac{I_{\text{c}}(0)}{t_{\text{junc}}W}$. Also, note that $\varphi(0)$ is independent of $y$ and therefore merely is a phase factor. The critical current is reached if we current-bias the junction by setting $\varphi(0) = \pi/2$, from which follows:

\begin{align} \label{Eq:H3_6}
\frac{I_{\text{c}}(B)}{I_{\text{c}}(0)} = \frac{1}{W} \left| \int_{-W/2}^{W/2} \cos \left( \frac{2\pi dB}{\Phi_0}y+2 \gamma\left(\frac{d}{2},y\right) \right) \mathrm{d}y \right|
\end{align}

\noindent We see that finding $I_{\text{c}}(B)$ becomes equal to a boundary condition problem of solving the Laplace equation in the geometry of the electrodes. Indeed, the solution is completely determined by the geometry of the sample and is independent of $\lambda$.

\section{Comparing different geometries}

As it is not trivial to find a general analytical solution to the boundary problem of Eq. \ref{Eq:H3_1} for the ellipsoid and rhomboid geometries, we solve the Laplace equation numerically using COMSOL Multiphysics 5.4. We define the right electrode geometry in 2D, divided into a triangular grid. Crucial for correctly solving Eq. \ref{Eq:H3_1}, is a grid size that is small enough to capture small changes in $\gamma$ and, on the edges, $\pmb{\hat{n}}_{\text{R}}$. We found a maximum element size (i.e., the grid edge size) of $0.01 \ln{(1+L/W)}$ nanometer to be a good compromise between computation time and precision. Using trigonometry we evaluate $\pmb{A}\cdot\pmb{\hat{n}}_{\text{R}}$ for each geometry and list the corresponding boundary conditions in Table \ref{table1} (here the numbering corresponds to the numbers in Figure \ref{H3_sim_geom}). In the Appendix, we provide a full derivation of each of the boundary conditions.

\begin{figure}[b!]
 \centerline{$
 \begin{array}{c}
  \includegraphics[width=0.85\linewidth]{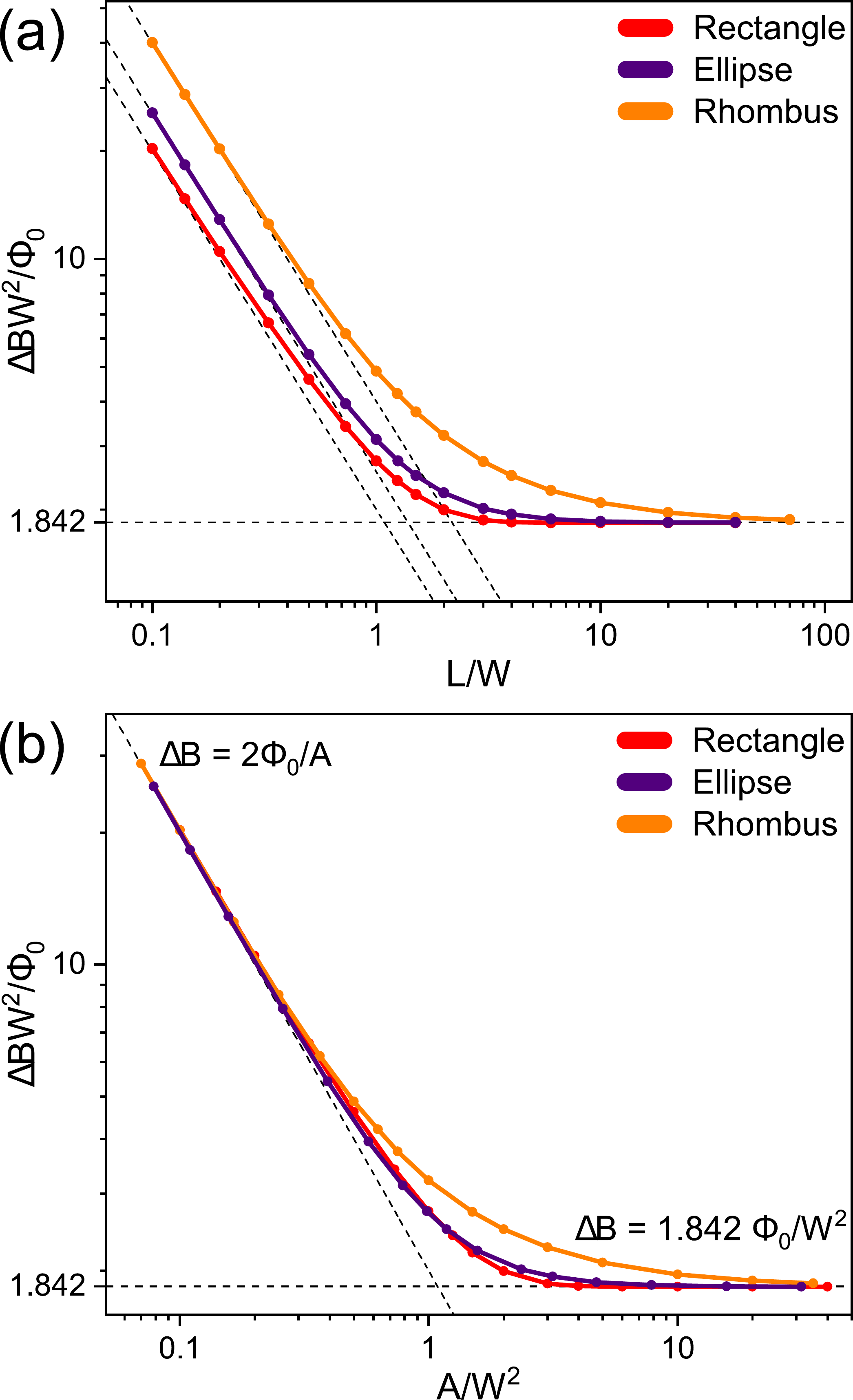}
 \end{array}$}
 \caption{Dimensionless measure of the period $\Delta B$ (the width of the fifth side lobe) of the calculated interference pattern $I_{\text{c}}(B)$ for the three geometries. In (a) we plot this value on log-log scale versus the aspect ratio $L/W$, in (b) it is plotted versus the total electrode area $A$ (i.e, combined area of left and right electrode), scaled by the $W^2$. Figure (b) reveals two limits for $\Delta B$ for $L \gg W$ and $L \ll W$. The first corresponds to the limit of an infinite superconducting strip $\Delta B = 1.842 \Phi_0 /W^2$, whereas in the latter we find $\Delta B = 2 \Phi_0 /A$. Contrary to $\Delta B$, $I_{\text{c}}(B)$ itself is not geometry independent in this limit.} \label{H3_DeltaB_results}
\end{figure}

\subsection{Simulation results}

Clem showed that the analytical solution for the rectangular geometry is an infinite series of sines and hyperbolic tangents\cite{Clem2010}. For the rectangle, this leads to the maximum in $\gamma \left(\frac{d}{2},y\right)$ to occur at $W/2$, which can be approximated as:

\begin{align} \label{Eq:H3_extra_FB}
\gamma \left(\frac{d}{2},\frac{W}{2}\right) = \frac{7~ \zeta(3) }{\pi^2} \frac{BW^2}{ \Phi_0}\tanh{\left(\frac{\pi^3}{28 ~ \zeta(3)} \frac{L}{W}\right)}
\end{align}

\noindent Here $\zeta$ is the Riemann zeta function. Now we generalize this approximation to include the other geometries. We find that the simulated $\gamma \left(\frac{d}{2},y\right)$ universally follows:

\begin{align} \label{Eq:H3_7}
\gamma \left(\frac{d}{2},y\right) = \frac{7 ~\zeta(3) }{\pi^2} \frac{BW^2}{ \Phi_0} \tanh{\left(\frac{\pi^3}{28~ \zeta(3)} \frac{A}{W^2}\right)} ~f\left(\frac{y}{W}\right)
\end{align}

\begin{figure}[t!]
 \centerline{$
 \begin{array}{c}
  \includegraphics[width=0.95\linewidth]{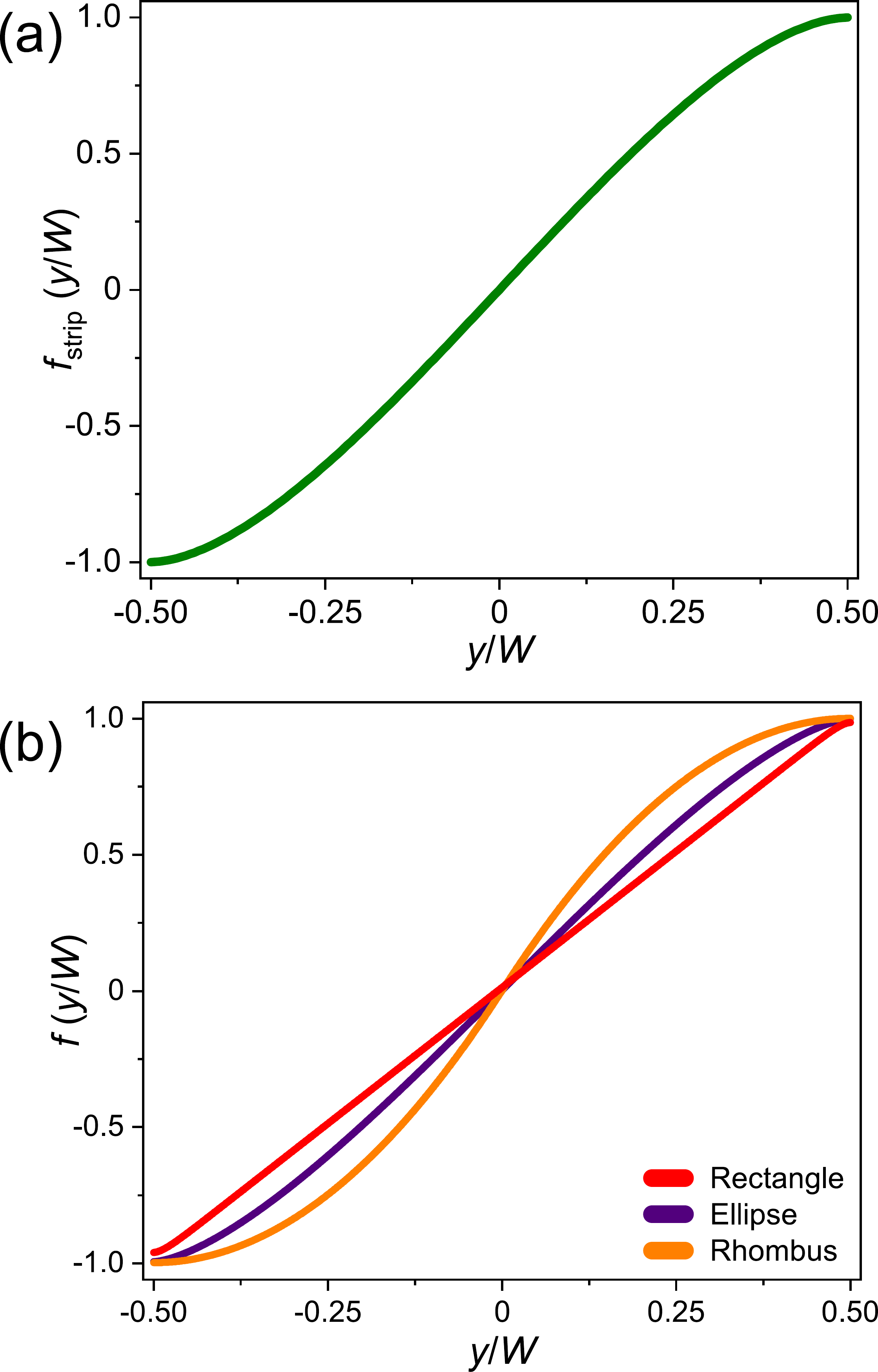}
 \end{array}$}
 \caption{dimensionless scaling functions $f\left(\frac{y}{W}\right)$ from Eq. \ref{Eq:H3_7}, for the limit $L \gg W$ in (a) and $L \ll W$ in (b). The maximum of these functions is located at $y = |W/2|$ and equals unity. Therefore, $\Delta B$ (large $n$ limit of the $n$th side lobe of $I_{\text{c}}(B)$) is universal for these limits. However, for the limit $L \ll W$, $f\left(\frac{y}{W}\right)$ is not geometry independent, which entails that $I_{\text{c}}(B)$ is not geometry independent as well, in this limit.}\label{H3_results_3}
\end{figure}

\noindent Where $f\left(\frac{y}{W}\right)$ is a dimensionless function defined by the specific geometry and $A$ is the total surface area of the electrodes (i.e, combined area of left and right electrode). Note that we have substituted $\frac{L}{W}$ in the argument of the hyperbolic tangent for $\frac{A}{W^2}$; the reason for this choice will become apparent below when discussing the period of the $I_{\text{c}}(B)$-pattern. Figure \ref{H3_disk_results}a shows the calculated $\gamma(x,y)$ for a disk geometry, normalized to the applied magnetic field and width of the electrodes $\gamma \Phi_0 / BW^2$. We plot $f\left(\frac{y}{W}\right)$ for this disk in Figure \ref{H3_disk_results}b. By evaluating the integral of Eq. \ref{Eq:H3_6} numerically for different values of $B$, we calculate the interference pattern of a disk-shaped junction (Figure \ref{H3_disk_results}c). the pattern resembles a Fraunhofer pattern at first sight. However, the peak height decreases less strongly than $1/B$, and the width of the middle lobe is not twice the width of the side lobes. In the inset of Figure \ref{H3_disk_results}c, we plot the width of the $n$th side lobe ($\Delta B_{\text{n}}$); the width increases and reaches an asymptotic value for large $n$.


In order to compare the interference patterns of junctions of different geometry, we define the period of the oscillations to be the width of the fifth side lobe ($\Delta B = \Delta B_5$). In the inset of Figure \ref{H3_disk_results}c, this is shown by the vertical reference line. The width of the fifth side lobe is not only sufficiently close to the asymptotic value but also experimentally accessible without the need for large magnetic fields. We now compare the periodicity of the interference patterns for different geometries by plotting the dimensionless value $\Delta B W^2 / \Phi_0$ as a function of the aspect ratio $L/W$ in Figure \ref{H3_DeltaB_results}a on a log-log scale. First, we find the results obtained on the rectangular junction to match the analytical results obtained by Clem\cite{Clem2010}. Furthermore, the periodicity of the pattern increases as the sample dimensions are diminished. Finally, we evaluated the width of the junction ($d$) to be irrelevant in determining $\Delta B$. Specifically, its contribution to the period is in the~{\textmu}T range for realistic sizes of $d$. The consequence is that $\Delta B$ is determined by the maximum of $\gamma$, i.e., $\gamma(\frac{d}{2},\frac{W}{2})$.


$\Delta B$ reaches asymptotic values for the limits $L \gg W$ and $L \ll W$ for all three geometries. The value of $\Delta B$ becomes geometry independent in these limits, as revealed by rescaling the results from Figure \ref{H3_DeltaB_results}a to a $\frac{A}{W^2}$ dependence, displayed in Figure \ref{H3_DeltaB_results}b. In the first limit, $L \gg W$, all three geometries become an infinite superconducting strip. Here we retrieve $\Delta B = 1.842 \Phi_0 /W^2$, which matches literature\cite{Moshe2008,Clem2010}. In this limit, we find $\gamma \left(\frac{d}{2},y\right)$ to follow:

\begin{align} \label{Eq:H3_extra1}
\gamma \left(\frac{d}{2},y\right) & = \frac{7 ~\zeta(3) }{\pi^2} \frac{BW^2}{ \Phi_0} f_{\text{strip}}\left(\frac{y}{W}\right) \\ & = \frac{\pi}{2} \frac{1}{1.842} \frac{BW^2}{ \Phi_0} f_{\text{strip}}\left(\frac{y}{W}\right)
\end{align}


\begin{figure*}[t!]
 \centerline{$
 \begin{array}{c}
  \includegraphics[width=0.8\linewidth]{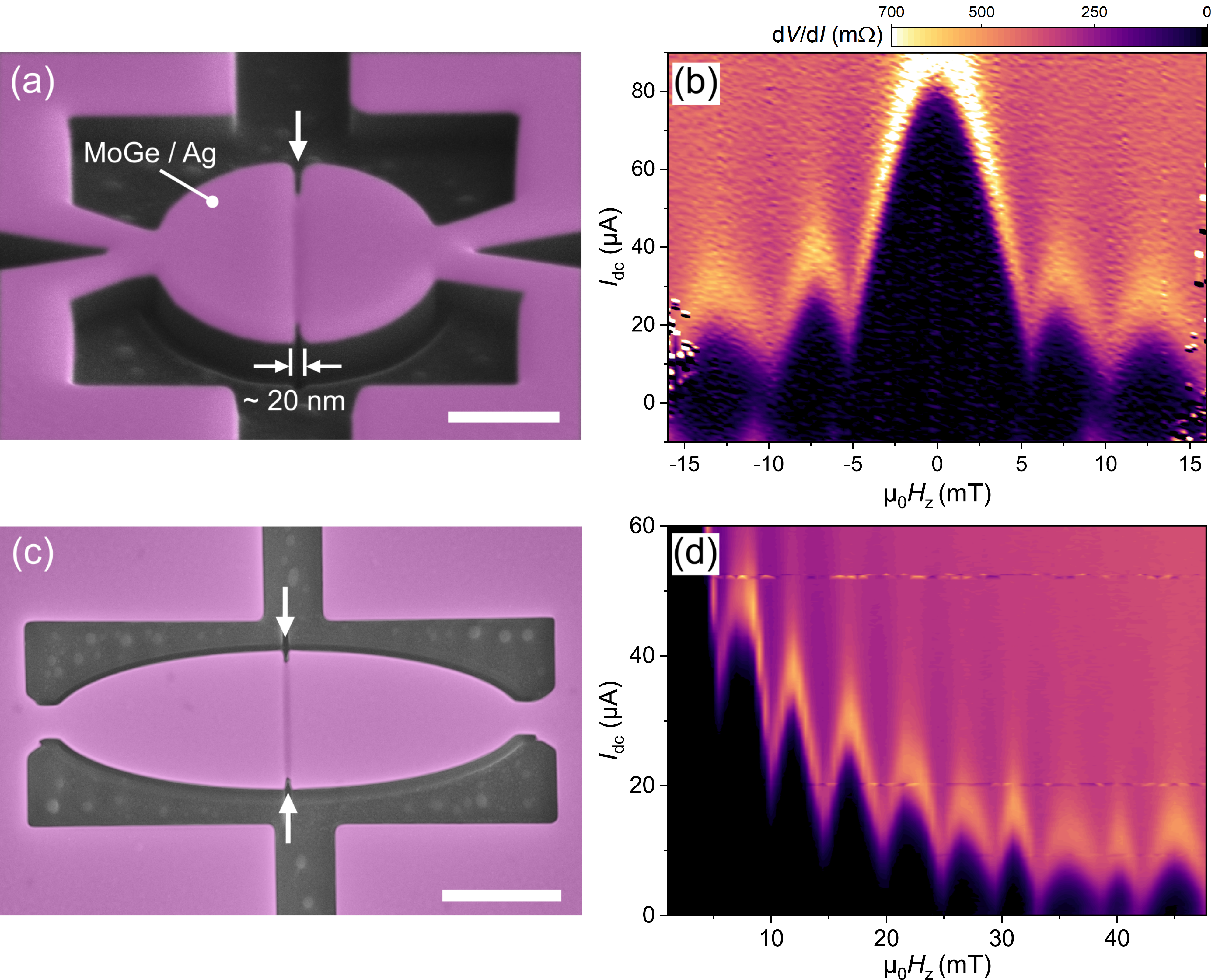}
 \end{array}$}
 \caption{Two S$-$N$-$S junction samples with a circular and ellipsoid geometry, produced from an Ag/MoGe bilayer and their corresponding $I_{\text{c}}(B)$-patterns, obtained at 2.5 K. (a) A false colored electron micrograph of a disk-shaped sample, viewed under an angle. The white arrow indicates the junction. Note the notches on the side of the sample due to an increased milling rate at the edges of the disk. The scale bar equals 500 nm. The corresponding $I_{\text{c}}(B)$ is displayed in (b) as a $dV/dI$ color map. As expected, the peak height of the side lobes is decreasing less rapidly than $1/B$. Contrary to the calculated pattern in Figure \ref{H3_disk_results}c, the middle peak is twice as wide as the neighboring ones. (c) depicts a top-view false colored electron micrograph of an ellipse-shaped junction. Again we indicate the notches with white arrows; the scale bar represents 1~{\textmu}m. In (d), we plot the corresponding interference pattern as a $dV/dI$ color map, which is used to extract the periodicity of the oscillations\cite{comment_paper}.}\label{H3_SEM_and_SQI}
\end{figure*}

\noindent Where $f_{\text{strip}}\left(\frac{y}{W}\right)$ is a dimensionless function running from -1 to 1, plotted in Figure \ref{H3_results_3}a. In the other limit, $L \ll W$, Eq. \ref{Eq:H3_7} reduces to:

\begin{align} \label{Eq:H3_extra2}
\gamma\left(\frac{d}{2},y\right) =  \frac{\pi A B}{4 \Phi_0 } f\left(\frac{y}{W}\right) = \frac{\pi}{2} \frac{ A B}{2 \Phi_0 } f\left(\frac{y}{W}\right)
\end{align}

Figure \ref{H3_results_3}b shows $f\left(\frac{y}{W}\right)$ in the limit $L \ll W$, for all three geometries. Since the maximum of $f\left(\frac{y}{W}\right)$ becomes independent of the underlying geometry and equal to unity, we find a geometry independent period, where $\Delta B = 2 \Phi_0 /A$. We can generalize this concept to find a general expression for $\Delta B$:

\begin{align} \label{Eq:H3_extra_FB2}
\Delta B = \frac{\pi}{2}\frac{1}{\max(\gamma/B)} = \frac{\pi}{2}\frac{B}{\gamma(\frac{d}{2},\frac{W}{2})}
\end{align}


\noindent Note that $\max(f\left(\frac{y}{W}\right)) \approx 1$ for all ratios $L/W$, and thus Eq. \ref{Eq:H3_extra_FB2} can serve as a good approximation for $\Delta B$. Therefore, we justify the relation of Eq. \ref{Eq:H3_7} as it demonstrates the emerging universal limits where $\Delta B = 2 \Phi_0 /A$ and $\Delta B = 1.842 \Phi_0 /W^2$, as well as provides a good approximation of $\Delta B$ between the limiting cases.

Although $\Delta B$ is geometry independent in the limit $L \ll W$, $I_{\text{c}}(B)$ itself is not universal in this limit. This is caused by the fact that $f\left(\frac{y}{W}\right)$ differs between geometries for $y \neq |W/2|$ (see Figure \ref{H3_results_3}b). For the rectangular geometry, for example, this function is linear in $y$: $f\left(\frac{y}{W}\right) = \frac{2y}{W}$. Therefore, we retrieve the Fraunhofer pattern, where $L_{\text{eff}} = L/2+d$. The effective length equals the length of a single superconducting electrode plus the junction length. This can be understood by considering that the screening currents trace loops in the electrodes, that reduce to two parallel and opposite current tracks, when $L \ll W$. $\gamma\left(\frac{d}{2},y\right)$ in the rhomboid geometry is radically different; it is well approximated by a sine function: $f\left(\frac{y}{W}\right) = \sin{\left(\frac{\pi y}{W}\right)}$. This leads to an interference pattern that is far closer to the pattern shown in Figure \ref{H3_disk_results}c, and not a Fraunhofer pattern. In conclusion: the shape and periodicity of the $I_{\text{c}}(B)$-pattern for low magnetic fields is independent of $\Delta B$, which is universal for $L \ll W$.


\subsection{Comparison to experiments}

In order to verify the dependence on the geometry, we fabricate five ellipse-shaped planar S$-$N$-$S junctions for different ratios of $L/W$. Besides, we make a rectangular-shaped junction with dimensions well in the $L \gg W$ limit.

First, a four-probe contact geometry is patterned on Si substrates using electron-beam lithography. Next, an Ag (20 nm), MoGe (55 nm) bilayer is deposited by sputter deposition. Subsequently, we use Focused Ion beam (FIB) milling to structure elliptical devices in the bilayer. By applying an ultra-low beam current of 1.5 pA, the weak link is formed by a line cut in the MoGe layer at the center of the device. This completely removes the superconductor on top, but leaves a normal metal connection. The resulting trench separates the MoGe electrodes by a roughly $20$ nm weak link, allowing Josephson coupling in this S$-$N$-$S system. Similar junctions, featuring a ferromagnetic layer, were fabricated in this manner, to study the interplay between supercurrents and ferromagnetic spin textures\cite{Lahabi2017a,Co_disk_paper,ellipse_paper2022}. Figures \ref{H3_SEM_and_SQI}a and \ref{H3_SEM_and_SQI}c show false colored electron micrographs of two of such devices, for $L=W$ and $L=4W$ respectively.

Two corresponding interference patterns obtained on the samples in \ref{H3_SEM_and_SQI}a and \ref{H3_SEM_and_SQI}c are shown in Figure \ref{H3_SEM_and_SQI}b and \ref{H3_SEM_and_SQI}d. Clearly, the period of the interference patterns scales with $L/W$. However, we find that the middle peak is twice the width of the neighboring ones and the amplitude of the side lobes of the $I_{\text{c}}(B)$-pattern feature a similar width, instead of the asymptotic behavior predicted by our theory (see Figure \ref{H3_disk_results}c). This can be explained by considering that $l \approx$ 100 nm (Eq. \ref{Eq:H3_JPL2}; based on $\lambda = 535$ nm\cite{Mandal2020}), which is small with respect to $W$. Our samples are therefore not in the narrow junction limit and allow Josephson vortices to stabilize in the junction. The width of the middle lobe can therefore not be predicted by our theory. However, Boris et al. have shown that $\Delta B_{\text{n}}$ for large $n$ follows the predictions of non-local electrodynamics\cite{Boris2013}. Therefore, we can compare the measured $\Delta B = \Delta B_5$ to our theoretical model.

\begin{figure}[t!]
 \centerline{$
 \begin{array}{c}
  \includegraphics[width=0.85\linewidth]{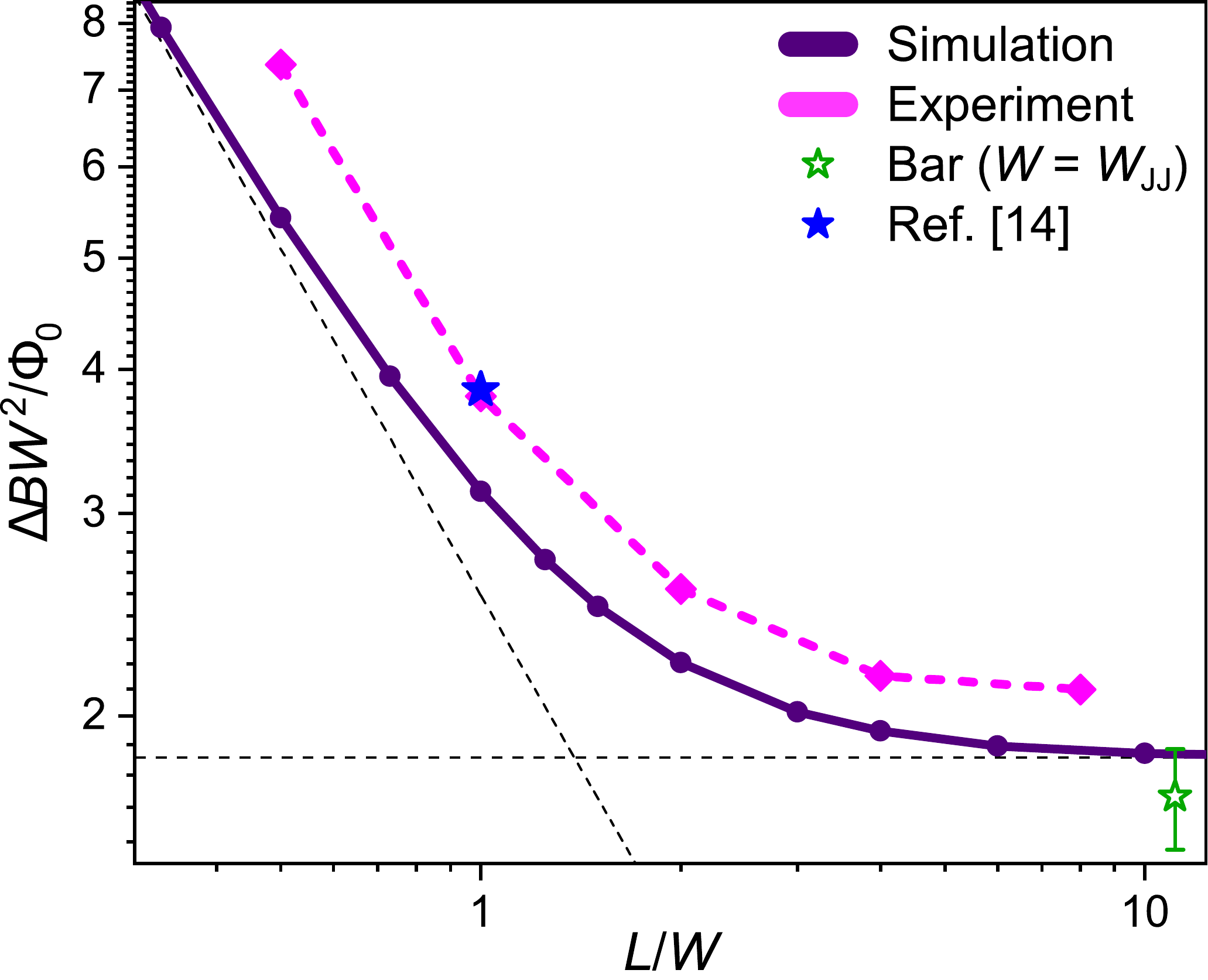}
 \end{array}$}
 \caption{Calculated periodicity $\Delta B$ of the fifth lobe of the interference pattern $I_{\text{c}}(B)$ obtained for the ellipse-shaped samples, compared to experimentally obtained values. We plot the dimensionless measure $\frac{\Delta B W^2}{\Phi_0}$ versus the aspect ratio $L/W$. The blue star indicates the periodicity of the cobalt-based disk junctions discussed in reference \cite{Co_disk_paper}. Although we can predicted the $L/W$-dependence, we find a constant offset between the experimental values and the simulations. This is due to the notches visible in Figure \ref{H3_SEM_and_SQI}a and \ref{H3_SEM_and_SQI}c, which makes the actual junction width ($W_{\text{JJ}}$) shorter than the width of the electrodes ($W$). To further illustrate this, we plot $\frac{\Delta B W_{\text{JJ}}^2}{\Phi_0}$ using the open green star.}\label{H3_DeltaB_experiments}
\end{figure} 

To compare the period of the $I_{\text{c}}(B)$-pattern to our theory, we plot $\Delta B$ for all measured samples along with the calculated values in Figure \ref{H3_DeltaB_experiments}. By the blue star symbol we also mark the periodicity of the Co-based S$-$F$-$S disk-junctions discussed elsewhere\cite{Co_disk_paper}. Although there is a constant offset between the measured periodicity and the calculated values, the overall trend is well predicted.

This constant offset is due to a trivial side effect of the FIB structuring method: some parts of the bi-layer (i.e., the edges of the device) mill faster than the bulk of the material. Consequently, notches develop on the side of the device when fabricating the trench. These notches make the width of the weak link ($W_{\text{JJ}}$) shorter than the width of the electrodes ($W$), which can result in a constant offset between experiments and the simulations, where it is assumed that $W_{\text{JJ}} = W$. In order to show that we reach the geometrical independent limit for $L \gg W$, we have fabricated a bar-shaped sample with $L/W>10$. In the Supplemental Material we present a scanning electron micrograph of this device accompanied by the interference pattern obtained on this sample\footnote{\label{note1}See the Supplemental Material for a description of the technical details of the Fourier transform and, results obtained on the bar-shaped sample.}. In this limit we expect $\Delta B W_{\text{JJ}}^2/\Phi_0 = 1.842$\cite{Golod2019,Hovhannisyan2022}. By inspection of the scanning electron micrograph we have extracted $W_{\text{JJ}}$ for the bar-shaped sample, which leads to $\Delta B W_{\text{JJ}}^2/\Phi_0 = 1.70$. By the green open star symbol, we plot $\Delta B W_{\text{JJ}}^2/\Phi_0$ for the bar-shaped sample in Figure \ref{H3_DeltaB_experiments}. The error bars correspond to a 20 nm uncertainty in the junction length.\footnote{We assume a relatively high uncertainty in the junction length due to the superconducting electrodes shielding the line of sight to the junction, which makes establishing $W_{\text{JJ}}$ more difficult. Furthermore, we have used $\Delta B = \Delta B_4$ since a rapid decay of the $I_{\text{c}}(B)$ pattern prevented establishing $\Delta B_5$. It must be noted that the average over the periodicity of the side peaks yields $\Delta B W_{\text{JJ}}^2/\Phi_0 = 1.83$.}

Another method of accounting for the influence of the notches is modifying the Fourier relation between the critical current density distribution $J(y)$ and the magnetic interference pattern $I_{\text{c}}(B)$, which will be discussed in the next section.

\section{Fourier analysis of thin film planar junctions.}

In their 1971 paper, Dynes and Fulton found a Fourier relation between the current density distribution of a Josephson junction and its magnetic interference pattern\cite{Dynes1971}. This method has been used widely the last years in analysing supercurrents planar Josephson junctions\cite{Hart2014,Pribiag2015,Huang2019,Suominen2017,Vries2018,Ying2020,Elfeky2021,Allen2016,Fortin2022,Lahabi2017a,Co_disk_paper}. However, the original Fourier relation is developed for macroscopic junctions where the screening currents are Meissner-based. This section will give a brief review of the Dynes and Fulton method and will adapt the Fourier relation for the use in thin film planar junctions, which is essential for correctly interpreting interference patterns obtained on such junctions.

First we write the current phase relation in Eq. \ref{Eq:21} as a complex expression and extend the integration bounds to infinity, since $J_{\text{c}}(y) = 0$, for $y > |W_{\text{JJ}}/2|$:

\begin{align} \label{Eq:23}
I(B) =  \operatorname{Im} \left( e^{i\varphi(0)}\int_{-\infty}^{\infty} J_{\text{c}}(y) e^{i\varphi_B} \: \text{d}y \right)
\end{align}

\noindent Here $\varphi_B$ is the gauge-invariant phase difference over the junction due to the magnetic induction. The critical current is given by the absolute value of the complex expression. Note that this equal to setting $\varphi(0) = \pi/2$ in Eq. \ref{Eq:H3_6}:

\begin{align} \label{Eq:24}
I_{\text{c}}(B) =   \left| \int_{-\infty}^{\infty} J_{\text{c}}(y) e^{i\varphi_B(B,y)} \: \text{d}y \right|
\end{align}

\noindent From this equation a general expression for a Fourier transform can be recognized. For a junction with macroscopic leads discussed above, we have $\varphi_B(B,y) = \frac{2\pi (2\lambda +d)B}{\Phi_0}y$ and therefore:

\begin{align} \label{Eq:25}
I_{\text{c}}(\beta) =   \left| \int_{-\infty}^{\infty} J_{\text{c}}(y) e^{2\pi i\beta y} \: \text{d}y \right|
\end{align}

\noindent Here we have defined the reduced field $\beta = \frac{(2\lambda +d)B}{\Phi_0}$, such that the position along the junctions $y$ and $\beta$ form conjugate variables. For the mesoscopic devices discussed here, this quantity needs to be replaced by Eq. \ref{Eq:H3_5}, yielding:

\begin{align} \label{Eq:H3_8}
I_{\text{c}}(B) =  \left| \int_{-\infty}^{\infty} J_{\text{c}}(y) e^{i2 \gamma\left(\frac{d}{2},y\right)} \: \text{d}y \right|
\end{align}

\noindent Where we omitted the contribution from the weak link, as its magnitude is negligible. Specifying $\gamma\left(\frac{d}{2},y\right)$ using Eq. \ref{Eq:H3_7}, we can define a new pair of conjugate variables: the length $\Tilde{y} = W f\left(\frac{y}{W}\right)$ and the reduced field $\Tilde{\beta} = \frac{7 ~\zeta(3) }{\pi^3} \frac{BW}{ \Phi_0} \tanh{\left(\frac{\pi^3}{28~ \zeta(3)} \frac{A}{W^2}\right)}$\footnote{Naturally, any choice of $\Tilde{y}$ and $\Tilde{\beta}$ is allowed, as long as it is consistent with $\gamma$.}, to arrive at:

\begin{align} \label{Eq:H3_9}
I_{\text{c}}(\Tilde{\beta}) =  \left| \int_{-\infty}^{\infty} \Tilde{J}_{\text{c}}(\Tilde{y}) e^{i2\pi \Tilde{\beta} \Tilde{y}} \: \text{d}\Tilde{y} \right|
\end{align}

\noindent Where we made a change of coordinates and $\Tilde{J}_{\text{c}}$ is defined as:

\begin{align} \label{Eq:H3_10}
\Tilde{J}_{\text{c}}\left(\frac{\Tilde{y}}{W}\right) = \frac{\text{d}g}{\text{d}\Tilde{y}}\left(\frac{\Tilde{y}}{W}\right) \: J_{\text{c}} \left( Wg\left(\frac{\Tilde{y}}{W}\right) \right) 
\end{align}

\noindent Here the function $g\left(\frac{\Tilde{y}}{W}\right)$ is the inverse of $f\left(\frac{y}{W}\right)$, or $g\left(\frac{\Tilde{y}}{W}\right) = f^{-1}\left(\frac{y}{W}\right)$.

\newpage

\begin{figure}[t!]
 \centerline{$
 \begin{array}{c}
  \includegraphics[width=0.85\linewidth]{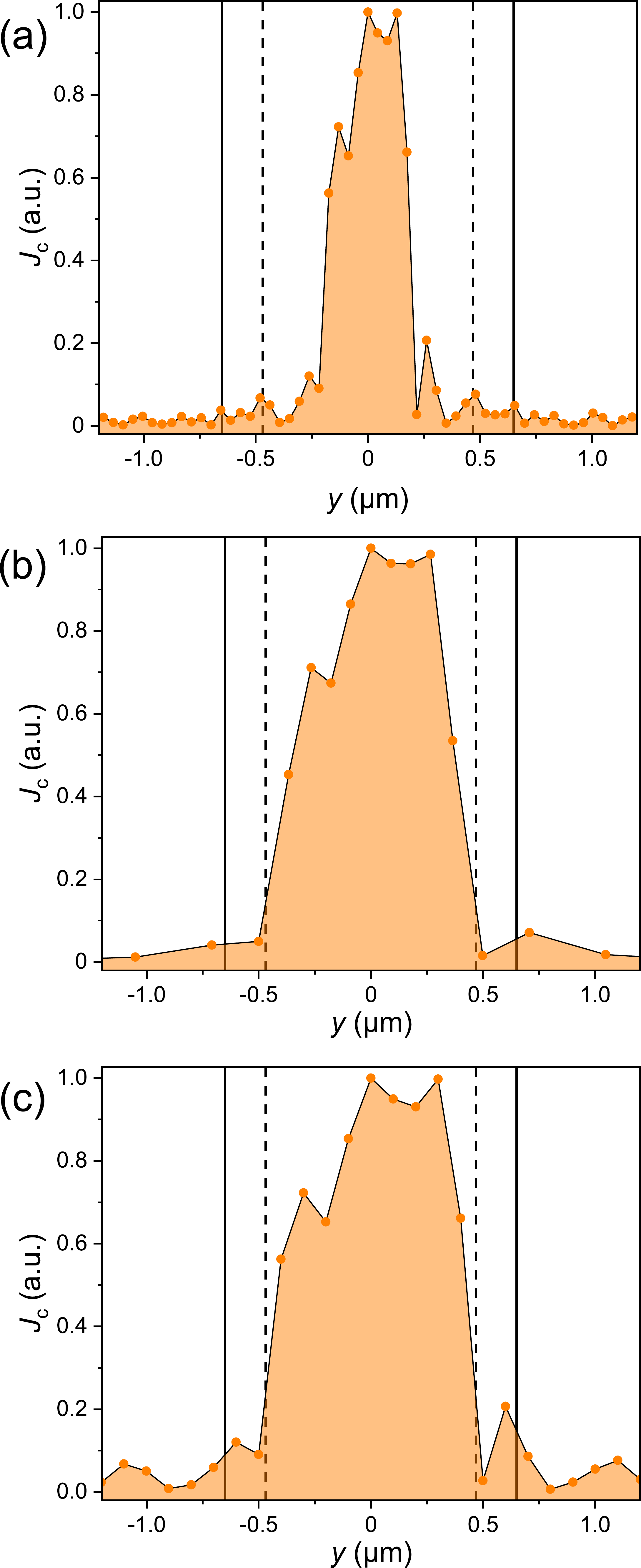}
 \end{array}$}
 \caption{Fourier analysis of the interference pattern shown in Figure \ref{H3_SEM_and_SQI}b, carried out using three different methods. In (a) we use the formalism for macroscopic junctions (following Eq. \ref{Eq:25}, where  $L_{\text{eff}} = 2\lambda+d$), whereas in (b), we make use of the simulation data shown in Figure \ref{H3_disk_results}b (following Eq. \ref{Eq:H3_8}). We indicate the boundaries of the electrodes ($-W/2$ and $W/2$) by solid reference lines and the boundaries of the actual weak link ($-W_{\text{JJ}}/2$ and $W_{\text{JJ}}/2$) by dotted reference lines. Only the method based on the simulations of the shielding currents correctly predicts the uniform current density distribution, which is limited to the actual junction only. Finally, in (c), we carried out the Fourier analysis using a linear approximation of $f_{\text{disk}}\left(\frac{y}{W}\right)$, circumventing the need for rescaling the axes, yet retaining the correct $J_{\text{c}}(y)$.} \label{H3_Fourier}
\end{figure} 

\clearpage

\noindent Equation \ref{Eq:H3_10} is a Fourier transform that includes a rescaling of the axes to retrieve the actual current density distribution $J_{\text{c}}(y)$. 

In Figure \ref{H3_Fourier} we compare three different methods of obtaining the current density distribution extracted by the Fourier analysis from the data obtained on the disk-shaped sample shown in Figure \ref{H3_SEM_and_SQI}b. Specifically, Figure \ref{H3_Fourier}a shows the current density distribution obtained using the method for macroscopic junctions (i.e., following Eq. \ref{Eq:25}, using $L_{\text{eff}} = 2\lambda+d$) and Figure \ref{H3_Fourier}b shows the Fourier transform based on our phase difference calculations (Eq. \ref{Eq:H3_8}). The solid reference lines indicate the width of the electrodes (i.e., the disk diameter $W$) and the dotted reference lines indicate the width of the actual junction as measured from the SEM micrograph ($W_{\text{JJ}}$).

\noindent We only observe a constant distribution of critical current throughout the full width of the junction (expected for uniform S$-$N$-$S junctions) when we incorporate the calculations presented in this paper. Contrarily, the analysis based on the $L_{\text{eff}} = 2\lambda+d$ yields an unphysical concentration of critical current in the middle of the junction. Finally, note that the current is confined to the actual junction ($W_{\text{JJ}}$), not the full width of the superconducting film ($W$). This explains the constant offset in Figure \ref{H3_DeltaB_experiments}a.

Alternatively, we can use a linear approximation of $\gamma\left(\frac{d}{2},y\right)$ to mitigate the need for rescaling the axes. Figure \ref{H3_Fourier}c shows the same Fourier analysis based on a linear approximation of $f\left(\frac{y}{W}\right)$. Since the linear approximation of $f_{\text{disk}}\left(\frac{y}{W}\right)$ breaks down near the edges, it yields less precise results at the junction boundaries. However, in the middle of the junction, the linear approximation of $f\left(\frac{y}{W}\right)$ is well suited for correctly analysing $J_{\text{c}}(y)$. For the technical details of carrying out the Fourier transform, the reader is referred to the Supplemental Material\cite{Note2}.

\section{Conclusion}

In conclusion, we analyzed the periodicity $\Delta B$ of the interference pattern $I_{\text{c}}(B)$ for thin film planar S$-$N$-$S Josephson junctions, both theoretically and experimentally. Specifically, we examine junctions separating rectangular, ellipsoid, and rhomboid films of width $W$ and length $L$. By mapping the second Ginzburg-Landau equation to the two-dimensional Laplace equation, we solve $I_{\text{c}}(B)$ for different ratios of $L/W$. We show that $\Delta B$ has two universal limits for $L \gg W$ and $L \ll W$, independent of the sample geometry. The first corresponds to an infinite superconducting strip, and the latter is caused by an emerging universal dependence of the phase difference on the junction electrode surface area. By fabricating elliptically-shaped S$-$N$-$S junctions, having different ratios for $L/W$, we experimentally verify the geometry dependence of $\Delta B$. Lastly, we adapt the Fourier relation between $I_{\text{c}}(B)$ and the critical current density distribution to suit planar junctions in the thin film limit. This proves to be vital in correctly predicting the location of current channels in topological planar Josephson junctions.

\section{Acknowledgements}
This work was supported by the Dutch Research Council (NWO) as part of the Frontiers of Nanoscience (NanoFront) program and through NWO projectruimte Grant No. 680.91.128. The work was also supported by EU Cost Action CA16218 (NANOCOHYBRI) and benefited from access to the Netherlands Centre for Electron Nanoscopy (NeCEN) at Leiden University. 

\appendix* 

\section{Derivation of the boundary conditions}

As discussed in the main text, the Neumann boundary conditions for the gauge-invariant phase are given by:

\begin{align} \label{APENDIX1}
(\pmb{\nabla}\gamma)\cdot\pmb{\hat{n}}_{\text{R}} = -\frac{2\pi}{\Phi_0}\pmb{A}\cdot\pmb{\hat{n}}_{\text{R}}
\end{align}

\noindent In this appendix, we will derive the results presented in Table \ref{table1}.

Combining the choice of the gauge $\pmb{A} = -yB\hat{x}$ with $\pmb{\hat{n}}_{\text{R}} = \hat{x}$ for boundary 1, we find:

\begin{align} \label{APENDIX2}
(\pmb{\nabla}\gamma)\cdot\pmb{\hat{n}}_{\text{R}} = (-\frac{2\pi}{\Phi_0}) (-yB) ~ \hat{x}\cdot\hat{x} = \frac{2\pi B}{\Phi_0} y
\end{align}

\noindent For boundary 2 we obtain the same result, yet with a minus sign since $\pmb{\hat{n}}_{\text{R}} = -\hat{x}$. For boundary 3 $\pmb{\hat{n}}_{\text{R}} = \pm \hat{y}$, which yields $(\pmb{\nabla}\gamma)\cdot\pmb{\hat{n}}_{\text{R}} \sim \hat{x}\cdot\hat{y} = 0$. 

Next, for boundary 4, parameterize the ellipse as $\frac{L}{2} \cos t ~\hat{x} + \frac{W}{2} \sin t ~ \hat{y}$. The tangent is then given by the derivative to $t$, which is $-\frac{L}{2} \sin t ~\hat{x} + \frac{W}{2} \cos t ~\hat{y} = -\frac{Ly}{W} ~ \hat{x} \frac{Wx}{L} ~\hat{y}$. Here, in the second step, we transformed back to Cartesian coordinates lying on the ellipse. A vector perpendicular to the tangent, pointing inwards to the ellipse, is then given by: $-\frac{Wx}{L}~ \hat{x} -\frac{Ly}{W} ~\hat{y}$. Normalizing yields $\pmb{\hat{n}}_{\text{R}}$:

\begin{align} \label{APENDIX3}
\pmb{\hat{n}}_{\text{R}} = -\frac{1}{\sqrt{\left(\frac{Wx}{L}\right)^2+\left(\frac{Ly}{W}\right)^2}} ~ (\frac{Wx}{L}\hat{x} + \frac{Ly}{W}\hat{y})
\end{align}

\noindent Taking the inner product with $\pmb{A}$, as in Eq. \ref{APENDIX2}, yields the boundary condition in Table \ref{table1}. Finally, for boundary 5, define the angle $\alpha$ as $\arctan (W/L)$. In that case, for $y>0$, we find $\pmb{\hat{n}}_{\text{R}} = -\sin \alpha ~ \hat{x} + -\cos \alpha ~ \hat{y}$, such that

\begin{align} \label{APENDIX4}
\pmb{\hat{n}}_{\text{R}} = -\frac{1}{\sqrt{W^2+L^2}} ~ (W \hat{x} + L \hat{y})
\end{align}

\noindent Again yielding the boundary condition in table \ref{table1}. Note that the boundary condition is unchanged for $y<0$, even though the $y$-component of $\pmb{\hat{n}}_{\text{R}}$ acquires a minus sign. This results from the choice of gauge ($A_y = 0$).

%


\begin{thebibliography}{34}%
\makeatletter
\providecommand \@ifxundefined [1]{%
 \@ifx{#1\undefined}
}%
\providecommand \@ifnum [1]{%
 \ifnum #1\expandafter \@firstoftwo
 \else \expandafter \@secondoftwo
 \fi
}%
\providecommand \@ifx [1]{%
 \ifx #1\expandafter \@firstoftwo
 \else \expandafter \@secondoftwo
 \fi
}%
\providecommand \natexlab [1]{#1}%
\providecommand \enquote  [1]{``#1''}%
\providecommand \bibnamefont  [1]{#1}%
\providecommand \bibfnamefont [1]{#1}%
\providecommand \citenamefont [1]{#1}%
\providecommand \href@noop [0]{\@secondoftwo}%
\providecommand \href [0]{\begingroup \@sanitize@url \@href}%
\providecommand \@href[1]{\@@startlink{#1}\@@href}%
\providecommand \@@href[1]{\endgroup#1\@@endlink}%
\providecommand \@sanitize@url [0]{\catcode `\\12\catcode `\$12\catcode
  `\&12\catcode `\#12\catcode `\^12\catcode `\_12\catcode `\%12\relax}%
\providecommand \@@startlink[1]{}%
\providecommand \@@endlink[0]{}%
\providecommand \url  [0]{\begingroup\@sanitize@url \@url }%
\providecommand \@url [1]{\endgroup\@href {#1}{\urlprefix }}%
\providecommand \urlprefix  [0]{URL }%
\providecommand \Eprint [0]{\href }%
\providecommand \doibase [0]{https://doi.org/}%
\providecommand \selectlanguage [0]{\@gobble}%
\providecommand \bibinfo  [0]{\@secondoftwo}%
\providecommand \bibfield  [0]{\@secondoftwo}%
\providecommand \translation [1]{[#1]}%
\providecommand \BibitemOpen [0]{}%
\providecommand \bibitemStop [0]{}%
\providecommand \bibitemNoStop [0]{.\EOS\space}%
\providecommand \EOS [0]{\spacefactor3000\relax}%
\providecommand \BibitemShut  [1]{\csname bibitem#1\endcsname}%
\let\auto@bib@innerbib\@empty
\bibitem [{\citenamefont {Hart}\ \emph {et~al.}(2014)\citenamefont {Hart},
  \citenamefont {Ren}, \citenamefont {Wagner}, \citenamefont {Leubner},
  \citenamefont {M{\"{u}}hlbauer}, \citenamefont {Br{\"{u}}ne}, \citenamefont
  {Buhmann}, \citenamefont {Molenkamp},\ and\ \citenamefont
  {Yacoby}}]{Hart2014}%
  \BibitemOpen
  \bibfield  {author} {\bibinfo {author} {\bibfnamefont {S.}~\bibnamefont
  {Hart}}, \bibinfo {author} {\bibfnamefont {H.}~\bibnamefont {Ren}}, \bibinfo
  {author} {\bibfnamefont {T.}~\bibnamefont {Wagner}}, \bibinfo {author}
  {\bibfnamefont {P.}~\bibnamefont {Leubner}}, \bibinfo {author} {\bibfnamefont
  {M.}~\bibnamefont {M{\"{u}}hlbauer}}, \bibinfo {author} {\bibfnamefont
  {C.}~\bibnamefont {Br{\"{u}}ne}}, \bibinfo {author} {\bibfnamefont
  {H.}~\bibnamefont {Buhmann}}, \bibinfo {author} {\bibfnamefont {L.~W.}\
  \bibnamefont {Molenkamp}},\ and\ \bibinfo {author} {\bibfnamefont
  {A.}~\bibnamefont {Yacoby}},\ }\bibfield  {title} {\bibinfo {title} {{Induced
  superconductivity in the quantum spin Hall edge}},\ }\href
  {https://doi.org/10.1038/nphys3036} {\bibfield  {journal} {\bibinfo
  {journal} {Nat. Phys.}\ }\textbf {\bibinfo {volume} {10}},\ \bibinfo {pages}
  {638} (\bibinfo {year} {2014})}\BibitemShut {NoStop}%
\bibitem [{\citenamefont {Pribiag}\ \emph {et~al.}(2015)\citenamefont
  {Pribiag}, \citenamefont {Beukman}, \citenamefont {Qu}, \citenamefont
  {Cassidy}, \citenamefont {Charpentier}, \citenamefont {Wegscheider},\ and\
  \citenamefont {Kouwenhoven}}]{Pribiag2015}%
  \BibitemOpen
  \bibfield  {author} {\bibinfo {author} {\bibfnamefont {V.~S.}\ \bibnamefont
  {Pribiag}}, \bibinfo {author} {\bibfnamefont {A.~J.~A.}\ \bibnamefont
  {Beukman}}, \bibinfo {author} {\bibfnamefont {F.}~\bibnamefont {Qu}},
  \bibinfo {author} {\bibfnamefont {M.~C.}\ \bibnamefont {Cassidy}}, \bibinfo
  {author} {\bibfnamefont {C.}~\bibnamefont {Charpentier}}, \bibinfo {author}
  {\bibfnamefont {W.}~\bibnamefont {Wegscheider}},\ and\ \bibinfo {author}
  {\bibfnamefont {L.~P.}\ \bibnamefont {Kouwenhoven}},\ }\bibfield  {title}
  {\bibinfo {title} {{Edge-mode superconductivity in a two-dimensional
  topological insulator}},\ }\href {https://doi.org/10.1038/nnano.2015.86}
  {\bibfield  {journal} {\bibinfo  {journal} {Nature Nanotechnology}\ }\textbf
  {\bibinfo {volume} {10}},\ \bibinfo {pages} {593} (\bibinfo {year}
  {2015})}\BibitemShut {NoStop}%
\bibitem [{\citenamefont {Fornieri}\ \emph {et~al.}(2019)\citenamefont
  {Fornieri}, \citenamefont {Whiticar}, \citenamefont {Setiawan}, \citenamefont
  {Portol{\'{e}}s}, \citenamefont {Drachmann}, \citenamefont {Keselman},
  \citenamefont {Gronin}, \citenamefont {Thomas}, \citenamefont {Wang},
  \citenamefont {Kallaher}, \citenamefont {Gardner}, \citenamefont {Berg},
  \citenamefont {Manfra}, \citenamefont {Stern}, \citenamefont {Marcus},\ and\
  \citenamefont {Nichele}}]{Fornieri2019}%
  \BibitemOpen
  \bibfield  {author} {\bibinfo {author} {\bibfnamefont {A.}~\bibnamefont
  {Fornieri}}, \bibinfo {author} {\bibfnamefont {A.~M.}\ \bibnamefont
  {Whiticar}}, \bibinfo {author} {\bibfnamefont {F.}~\bibnamefont {Setiawan}},
  \bibinfo {author} {\bibfnamefont {E.}~\bibnamefont {Portol{\'{e}}s}},
  \bibinfo {author} {\bibfnamefont {A.~C.~C.}\ \bibnamefont {Drachmann}},
  \bibinfo {author} {\bibfnamefont {A.}~\bibnamefont {Keselman}}, \bibinfo
  {author} {\bibfnamefont {S.}~\bibnamefont {Gronin}}, \bibinfo {author}
  {\bibfnamefont {C.}~\bibnamefont {Thomas}}, \bibinfo {author} {\bibfnamefont
  {T.}~\bibnamefont {Wang}}, \bibinfo {author} {\bibfnamefont {R.}~\bibnamefont
  {Kallaher}}, \bibinfo {author} {\bibfnamefont {G.~C.}\ \bibnamefont
  {Gardner}}, \bibinfo {author} {\bibfnamefont {E.}~\bibnamefont {Berg}},
  \bibinfo {author} {\bibfnamefont {M.~J.}\ \bibnamefont {Manfra}}, \bibinfo
  {author} {\bibfnamefont {A.}~\bibnamefont {Stern}}, \bibinfo {author}
  {\bibfnamefont {C.~M.}\ \bibnamefont {Marcus}},\ and\ \bibinfo {author}
  {\bibfnamefont {F.}~\bibnamefont {Nichele}},\ }\bibfield  {title} {\bibinfo
  {title} {{Evidence of topological superconductivity in planar Josephson
  junctions}},\ }\href {https://doi.org/10.1038/s41586-019-1068-8} {\bibfield
  {journal} {\bibinfo  {journal} {Nature}\ }\textbf {\bibinfo {volume} {569}},\
  \bibinfo {pages} {89} (\bibinfo {year} {2019})}\BibitemShut {NoStop}%
\bibitem [{\citenamefont {Mayer}\ \emph {et~al.}(1993)\citenamefont {Mayer},
  \citenamefont {Schuster}, \citenamefont {Beck}, \citenamefont {Alff},\ and\
  \citenamefont {Gross}}]{Mayer1993}%
  \BibitemOpen
  \bibfield  {author} {\bibinfo {author} {\bibfnamefont {B.}~\bibnamefont
  {Mayer}}, \bibinfo {author} {\bibfnamefont {S.}~\bibnamefont {Schuster}},
  \bibinfo {author} {\bibfnamefont {A.}~\bibnamefont {Beck}}, \bibinfo {author}
  {\bibfnamefont {L.}~\bibnamefont {Alff}},\ and\ \bibinfo {author}
  {\bibfnamefont {R.}~\bibnamefont {Gross}},\ }\bibfield  {title} {\bibinfo
  {title} {{Magnetic field dependence of the critical current in
  YBa$_2$Cu$_3$O$_{7-\delta}$ bicrystal grain boundary junctions}},\ }\href
  {https://doi.org/10.1063/1.108578} {\bibfield  {journal} {\bibinfo  {journal}
  {Applied Physics Letters}\ }\textbf {\bibinfo {volume} {62}},\ \bibinfo
  {pages} {783} (\bibinfo {year} {1993})}\BibitemShut {NoStop}%
\bibitem [{\citenamefont {Cybart}\ \emph {et~al.}(2015)\citenamefont {Cybart},
  \citenamefont {Cho}, \citenamefont {Wong}, \citenamefont {Wehlin},
  \citenamefont {Ma}, \citenamefont {Huynh},\ and\ \citenamefont
  {Dynes}}]{Cybart2015a}%
  \BibitemOpen
  \bibfield  {author} {\bibinfo {author} {\bibfnamefont {S.~A.}\ \bibnamefont
  {Cybart}}, \bibinfo {author} {\bibfnamefont {E.~Y.}\ \bibnamefont {Cho}},
  \bibinfo {author} {\bibfnamefont {T.~J.}\ \bibnamefont {Wong}}, \bibinfo
  {author} {\bibfnamefont {B.~H.}\ \bibnamefont {Wehlin}}, \bibinfo {author}
  {\bibfnamefont {M.~K.}\ \bibnamefont {Ma}}, \bibinfo {author} {\bibfnamefont
  {C.}~\bibnamefont {Huynh}},\ and\ \bibinfo {author} {\bibfnamefont {R.~C.}\
  \bibnamefont {Dynes}},\ }\bibfield  {title} {\bibinfo {title} {{Nano
  Josephson superconducting tunnel junctions in YBa$_2$Cu$_3$O$_{7-\delta}$
  directly patterned with a focused helium ion beam}},\ }\href
  {https://doi.org/10.1038/nnano.2015.76} {\bibfield  {journal} {\bibinfo
  {journal} {Nature Nanotechnology}\ }\textbf {\bibinfo {volume} {10}},\
  \bibinfo {pages} {598} (\bibinfo {year} {2015})}\BibitemShut {NoStop}%
\bibitem [{\citenamefont {Ying}\ \emph {et~al.}(2020)\citenamefont {Ying},
  \citenamefont {He}, \citenamefont {Yang}, \citenamefont {Liu}, \citenamefont
  {Lyu}, \citenamefont {Zhang}, \citenamefont {Liu}, \citenamefont {Zhao},
  \citenamefont {Jiang}, \citenamefont {Ji}, \citenamefont {Fan}, \citenamefont
  {Yang}, \citenamefont {Jing}, \citenamefont {Liu}, \citenamefont {Cao},
  \citenamefont {Wang}, \citenamefont {Lu},\ and\ \citenamefont
  {Qu}}]{Ying2020}%
  \BibitemOpen
  \bibfield  {author} {\bibinfo {author} {\bibfnamefont {J.}~\bibnamefont
  {Ying}}, \bibinfo {author} {\bibfnamefont {J.}~\bibnamefont {He}}, \bibinfo
  {author} {\bibfnamefont {G.}~\bibnamefont {Yang}}, \bibinfo {author}
  {\bibfnamefont {M.}~\bibnamefont {Liu}}, \bibinfo {author} {\bibfnamefont
  {Z.}~\bibnamefont {Lyu}}, \bibinfo {author} {\bibfnamefont {X.}~\bibnamefont
  {Zhang}}, \bibinfo {author} {\bibfnamefont {H.}~\bibnamefont {Liu}}, \bibinfo
  {author} {\bibfnamefont {K.}~\bibnamefont {Zhao}}, \bibinfo {author}
  {\bibfnamefont {R.}~\bibnamefont {Jiang}}, \bibinfo {author} {\bibfnamefont
  {Z.}~\bibnamefont {Ji}}, \bibinfo {author} {\bibfnamefont {J.}~\bibnamefont
  {Fan}}, \bibinfo {author} {\bibfnamefont {C.}~\bibnamefont {Yang}}, \bibinfo
  {author} {\bibfnamefont {X.}~\bibnamefont {Jing}}, \bibinfo {author}
  {\bibfnamefont {G.}~\bibnamefont {Liu}}, \bibinfo {author} {\bibfnamefont
  {X.}~\bibnamefont {Cao}}, \bibinfo {author} {\bibfnamefont {X.}~\bibnamefont
  {Wang}}, \bibinfo {author} {\bibfnamefont {L.}~\bibnamefont {Lu}},\ and\
  \bibinfo {author} {\bibfnamefont {F.}~\bibnamefont {Qu}},\ }\bibfield
  {title} {\bibinfo {title} {{Magnitude and spatial distribution control of the
  supercurrent in Bi$_2$O$_2$Se-based Josephson junction}},\ }\href
  {https://doi.org/10.1021/acs.nanolett.0c00025} {\bibfield  {journal}
  {\bibinfo  {journal} {Nano Letters}\ }\textbf {\bibinfo {volume} {20}},\
  \bibinfo {pages} {2569} (\bibinfo {year} {2020})}\BibitemShut {NoStop}%
\bibitem [{\citenamefont {Elfeky}\ \emph {et~al.}(2021)\citenamefont {Elfeky},
  \citenamefont {Lotfizadeh}, \citenamefont {Schiela}, \citenamefont
  {Strickland}, \citenamefont {Dartiailh}, \citenamefont {Sardashti},
  \citenamefont {Hatefipour}, \citenamefont {Yu}, \citenamefont {Pankratova},
  \citenamefont {Lee}, \citenamefont {Manucharyan},\ and\ \citenamefont
  {Shabani}}]{Elfeky2021}%
  \BibitemOpen
  \bibfield  {author} {\bibinfo {author} {\bibfnamefont {B.~H.}\ \bibnamefont
  {Elfeky}}, \bibinfo {author} {\bibfnamefont {N.}~\bibnamefont {Lotfizadeh}},
  \bibinfo {author} {\bibfnamefont {W.~F.}\ \bibnamefont {Schiela}}, \bibinfo
  {author} {\bibfnamefont {W.~M.}\ \bibnamefont {Strickland}}, \bibinfo
  {author} {\bibfnamefont {M.}~\bibnamefont {Dartiailh}}, \bibinfo {author}
  {\bibfnamefont {K.}~\bibnamefont {Sardashti}}, \bibinfo {author}
  {\bibfnamefont {M.}~\bibnamefont {Hatefipour}}, \bibinfo {author}
  {\bibfnamefont {P.}~\bibnamefont {Yu}}, \bibinfo {author} {\bibfnamefont
  {N.}~\bibnamefont {Pankratova}}, \bibinfo {author} {\bibfnamefont
  {H.}~\bibnamefont {Lee}}, \bibinfo {author} {\bibfnamefont {V.~E.}\
  \bibnamefont {Manucharyan}},\ and\ \bibinfo {author} {\bibfnamefont
  {J.}~\bibnamefont {Shabani}},\ }\bibfield  {title} {\bibinfo {title} {{Local
  control of supercurrent density in epitaxial planar Josephson junctions}},\
  }\href {https://doi.org/10.1021/acs.nanolett.1c02771} {\bibfield  {journal}
  {\bibinfo  {journal} {Nano Letters}\ }\textbf {\bibinfo {volume} {21}},\
  \bibinfo {pages} {8274} (\bibinfo {year} {2021})}\BibitemShut {NoStop}%
\bibitem [{\citenamefont {Allen}\ \emph {et~al.}(2016)\citenamefont {Allen},
  \citenamefont {Shtanko}, \citenamefont {Fulga}, \citenamefont {Akhmerov},
  \citenamefont {Watanabe}, \citenamefont {Taniguchi}, \citenamefont
  {Jarillo-Herrero}, \citenamefont {Levitov},\ and\ \citenamefont
  {Yacoby}}]{Allen2016}%
  \BibitemOpen
  \bibfield  {author} {\bibinfo {author} {\bibfnamefont {M.~T.}\ \bibnamefont
  {Allen}}, \bibinfo {author} {\bibfnamefont {O.}~\bibnamefont {Shtanko}},
  \bibinfo {author} {\bibfnamefont {I.~C.}\ \bibnamefont {Fulga}}, \bibinfo
  {author} {\bibfnamefont {A.~R.}\ \bibnamefont {Akhmerov}}, \bibinfo {author}
  {\bibfnamefont {K.}~\bibnamefont {Watanabe}}, \bibinfo {author}
  {\bibfnamefont {T.}~\bibnamefont {Taniguchi}}, \bibinfo {author}
  {\bibfnamefont {P.}~\bibnamefont {Jarillo-Herrero}}, \bibinfo {author}
  {\bibfnamefont {L.~S.}\ \bibnamefont {Levitov}},\ and\ \bibinfo {author}
  {\bibfnamefont {A.}~\bibnamefont {Yacoby}},\ }\bibfield  {title} {\bibinfo
  {title} {{Spatially resolved edge currents and guided-wave electronic states
  in graphene}},\ }\href {https://doi.org/10.1038/nphys3534} {\bibfield
  {journal} {\bibinfo  {journal} {Nature Physics}\ }\textbf {\bibinfo {volume}
  {12}},\ \bibinfo {pages} {128} (\bibinfo {year} {2016})}\BibitemShut
  {NoStop}%
\bibitem [{\citenamefont {Fortin-Desch{\^{e}}nes}\ \emph
  {et~al.}(2022)\citenamefont {Fortin-Desch{\^{e}}nes}, \citenamefont {Pu},
  \citenamefont {Zhou}, \citenamefont {Ma}, \citenamefont {Cheung},
  \citenamefont {Watanabe}, \citenamefont {Taniguchi}, \citenamefont {Zhang},
  \citenamefont {Du},\ and\ \citenamefont {Xia}}]{Fortin2022}%
  \BibitemOpen
  \bibfield  {author} {\bibinfo {author} {\bibfnamefont {M.}~\bibnamefont
  {Fortin-Desch{\^{e}}nes}}, \bibinfo {author} {\bibfnamefont {R.}~\bibnamefont
  {Pu}}, \bibinfo {author} {\bibfnamefont {Y.-f.}\ \bibnamefont {Zhou}},
  \bibinfo {author} {\bibfnamefont {C.}~\bibnamefont {Ma}}, \bibinfo {author}
  {\bibfnamefont {P.}~\bibnamefont {Cheung}}, \bibinfo {author} {\bibfnamefont
  {K.}~\bibnamefont {Watanabe}}, \bibinfo {author} {\bibfnamefont
  {T.}~\bibnamefont {Taniguchi}}, \bibinfo {author} {\bibfnamefont
  {F.}~\bibnamefont {Zhang}}, \bibinfo {author} {\bibfnamefont
  {X.}~\bibnamefont {Du}},\ and\ \bibinfo {author} {\bibfnamefont
  {F.}~\bibnamefont {Xia}},\ }\bibfield  {title} {\bibinfo {title} {{Uncovering
  topological edge states in twisted bilayer graphene}},\ }\href
  {https://doi.org/10.1021/acs.nanolett.2c01481} {\bibfield  {journal}
  {\bibinfo  {journal} {Nano Letters}\ }\textbf {\bibinfo {volume} {22}},\
  \bibinfo {pages} {6186} (\bibinfo {year} {2022})}\BibitemShut {NoStop}%
\bibitem [{\citenamefont {Golod}\ \emph {et~al.}(2019)\citenamefont {Golod},
  \citenamefont {Kapran},\ and\ \citenamefont {Krasnov}}]{Golod2019}%
  \BibitemOpen
  \bibfield  {author} {\bibinfo {author} {\bibfnamefont {T.}~\bibnamefont
  {Golod}}, \bibinfo {author} {\bibfnamefont {O.~M.}\ \bibnamefont {Kapran}},\
  and\ \bibinfo {author} {\bibfnamefont {V.~M.}\ \bibnamefont {Krasnov}},\
  }\bibfield  {title} {\bibinfo {title} {{Planar
  superconductor-ferromagnet-superconductor Josephson junctions as
  scanning-probe sensors}},\ }\href
  {https://doi.org/10.1103/PhysRevApplied.11.014062} {\bibfield  {journal}
  {\bibinfo  {journal} {Phys. Rev. Applied}\ }\textbf {\bibinfo {volume}
  {11}},\ \bibinfo {pages} {014062} (\bibinfo {year} {2019})}\BibitemShut
  {NoStop}%
\bibitem [{\citenamefont {LeFebvre}\ \emph {et~al.}(2022)\citenamefont
  {LeFebvre}, \citenamefont {Cho}, \citenamefont {Li}, \citenamefont {Cai},\
  and\ \citenamefont {Cybart}}]{LeFebvre2022}%
  \BibitemOpen
  \bibfield  {author} {\bibinfo {author} {\bibfnamefont {J.~C.}\ \bibnamefont
  {LeFebvre}}, \bibinfo {author} {\bibfnamefont {E.}~\bibnamefont {Cho}},
  \bibinfo {author} {\bibfnamefont {H.}~\bibnamefont {Li}}, \bibinfo {author}
  {\bibfnamefont {H.}~\bibnamefont {Cai}},\ and\ \bibinfo {author}
  {\bibfnamefont {S.~A.}\ \bibnamefont {Cybart}},\ }\bibfield  {title}
  {\bibinfo {title} {{Flux focused series arrays of long Josephson junctions
  for high-dynamic range magnetic field sensing}},\ }\href
  {https://doi.org/10.1063/5.0087611} {\bibfield  {journal} {\bibinfo
  {journal} {Journal of Applied Physics}\ }\textbf {\bibinfo {volume} {131}},\
  \bibinfo {pages} {163902} (\bibinfo {year} {2022})}\BibitemShut {NoStop}%
\bibitem [{\citenamefont {Hovhannisyan}\ \emph {et~al.}(2022)\citenamefont
  {Hovhannisyan}, \citenamefont {Golod},\ and\ \citenamefont
  {Krasnov}}]{Hovhannisyan2022}%
  \BibitemOpen
  \bibfield  {author} {\bibinfo {author} {\bibfnamefont {R.~A.}\ \bibnamefont
  {Hovhannisyan}}, \bibinfo {author} {\bibfnamefont {T.}~\bibnamefont
  {Golod}},\ and\ \bibinfo {author} {\bibfnamefont {V.~M.}\ \bibnamefont
  {Krasnov}},\ }\bibfield  {title} {\bibinfo {title} {{Holographic
  reconstruction of magnetic field distribution in a Josephson junction from
  diffraction-like ${I}_{c}(H)$ patterns}},\ }\href
  {https://doi.org/10.1103/PhysRevB.105.214513} {\bibfield  {journal} {\bibinfo
   {journal} {Phys. Rev. B}\ }\textbf {\bibinfo {volume} {105}},\ \bibinfo
  {pages} {214513} (\bibinfo {year} {2022})}\BibitemShut {NoStop}%
\bibitem [{\citenamefont {Lahabi}\ \emph {et~al.}(2017)\citenamefont {Lahabi},
  \citenamefont {Amundsen}, \citenamefont {Ouassou}, \citenamefont {Beukers},
  \citenamefont {Pleijster}, \citenamefont {Linder}, \citenamefont {Alkemade},\
  and\ \citenamefont {Aarts}}]{Lahabi2017a}%
  \BibitemOpen
  \bibfield  {author} {\bibinfo {author} {\bibfnamefont {K.}~\bibnamefont
  {Lahabi}}, \bibinfo {author} {\bibfnamefont {M.}~\bibnamefont {Amundsen}},
  \bibinfo {author} {\bibfnamefont {J.~A.}\ \bibnamefont {Ouassou}}, \bibinfo
  {author} {\bibfnamefont {E.}~\bibnamefont {Beukers}}, \bibinfo {author}
  {\bibfnamefont {M.}~\bibnamefont {Pleijster}}, \bibinfo {author}
  {\bibfnamefont {J.}~\bibnamefont {Linder}}, \bibinfo {author} {\bibfnamefont
  {P.}~\bibnamefont {Alkemade}},\ and\ \bibinfo {author} {\bibfnamefont
  {J.}~\bibnamefont {Aarts}},\ }\bibfield  {title} {\bibinfo {title}
  {{Controlling supercurrents and their spatial distribution in
  ferromagnets}},\ }\href {https://doi.org/10.1038/s41467-017-02236-2}
  {\bibfield  {journal} {\bibinfo  {journal} {Nat. Commun.}\ }\textbf {\bibinfo
  {volume} {8}},\ \bibinfo {pages} {2056} (\bibinfo {year} {2017})}\BibitemShut
  {NoStop}%
\bibitem [{\citenamefont {Fermin}\ \emph
  {et~al.}(2022{\natexlab{a}})\citenamefont {Fermin}, \citenamefont {van
  Dinter}, \citenamefont {Hubert}, \citenamefont {Woltjes}, \citenamefont
  {Silaev}, \citenamefont {Aarts},\ and\ \citenamefont
  {Lahabi}}]{Co_disk_paper}%
  \BibitemOpen
  \bibfield  {author} {\bibinfo {author} {\bibfnamefont {R.}~\bibnamefont
  {Fermin}}, \bibinfo {author} {\bibfnamefont {D.}~\bibnamefont {van Dinter}},
  \bibinfo {author} {\bibfnamefont {M.}~\bibnamefont {Hubert}}, \bibinfo
  {author} {\bibfnamefont {B.}~\bibnamefont {Woltjes}}, \bibinfo {author}
  {\bibfnamefont {M.}~\bibnamefont {Silaev}}, \bibinfo {author} {\bibfnamefont
  {J.}~\bibnamefont {Aarts}},\ and\ \bibinfo {author} {\bibfnamefont
  {K.}~\bibnamefont {Lahabi}},\ }\bibfield  {title} {\bibinfo {title}
  {Superconducting triplet rim currents in a spin-textured ferromagnetic
  disk},\ }\href {https://doi.org/10.1021/acs.nanolett.1c04051} {\bibfield
  {journal} {\bibinfo  {journal} {Nano Letters}\ }\textbf {\bibinfo {volume}
  {22}},\ \bibinfo {pages} {2209} (\bibinfo {year}
  {2022}{\natexlab{a}})}\BibitemShut {NoStop}%
\bibitem [{\citenamefont {Jeon}\ \emph {et~al.}(2021)\citenamefont {Jeon},
  \citenamefont {Hazra}, \citenamefont {Cho}, \citenamefont {Chakraborty},
  \citenamefont {Jeon}, \citenamefont {Han}, \citenamefont {Meyerheim},
  \citenamefont {Kontos},\ and\ \citenamefont {Parkin}}]{Jeon2021}%
  \BibitemOpen
  \bibfield  {author} {\bibinfo {author} {\bibfnamefont {K.~R.}\ \bibnamefont
  {Jeon}}, \bibinfo {author} {\bibfnamefont {B.~K.}\ \bibnamefont {Hazra}},
  \bibinfo {author} {\bibfnamefont {K.}~\bibnamefont {Cho}}, \bibinfo {author}
  {\bibfnamefont {A.}~\bibnamefont {Chakraborty}}, \bibinfo {author}
  {\bibfnamefont {J.~C.}\ \bibnamefont {Jeon}}, \bibinfo {author}
  {\bibfnamefont {H.}~\bibnamefont {Han}}, \bibinfo {author} {\bibfnamefont
  {H.~L.}\ \bibnamefont {Meyerheim}}, \bibinfo {author} {\bibfnamefont
  {T.}~\bibnamefont {Kontos}},\ and\ \bibinfo {author} {\bibfnamefont {S.~S.}\
  \bibnamefont {Parkin}},\ }\bibfield  {title} {\bibinfo {title} {{Long-range
  supercurrents through a chiral non-collinear antiferromagnet in lateral
  Josephson junctions}},\ }\href {https://doi.org/10.1038/s41563-021-01061-9}
  {\bibfield  {journal} {\bibinfo  {journal} {Nature Materials}\ }\textbf
  {\bibinfo {volume} {20}},\ \bibinfo {pages} {1358} (\bibinfo {year}
  {2021})}\BibitemShut {NoStop}%
\bibitem [{\citenamefont {Dynes}\ and\ \citenamefont
  {Fulton}(1971)}]{Dynes1971}%
  \BibitemOpen
  \bibfield  {author} {\bibinfo {author} {\bibfnamefont {R.~C.}\ \bibnamefont
  {Dynes}}\ and\ \bibinfo {author} {\bibfnamefont {T.~A.}\ \bibnamefont
  {Fulton}},\ }\bibfield  {title} {\bibinfo {title} {{Supercurrent density
  distribution in Josephson junctions}},\ }\href
  {https://doi.org/10.1103/PhysRevB.3.3015} {\bibfield  {journal} {\bibinfo
  {journal} {Phys. Rev. B}\ }\textbf {\bibinfo {volume} {3}},\ \bibinfo {pages}
  {3015} (\bibinfo {year} {1971})}\BibitemShut {NoStop}%
\bibitem [{\citenamefont {Pearl}(1964)}]{Pearl1964}%
  \BibitemOpen
  \bibfield  {author} {\bibinfo {author} {\bibfnamefont {J.}~\bibnamefont
  {Pearl}},\ }\bibfield  {title} {\bibinfo {title} {Current distribution in
  superconducting films carrying quantized fluxoids},\ }\href
  {https://doi.org/10.1063/1.1754056} {\bibfield  {journal} {\bibinfo
  {journal} {Appl. Phys. Lett.}\ }\textbf {\bibinfo {volume} {5}},\ \bibinfo
  {pages} {65} (\bibinfo {year} {1964})}\BibitemShut {NoStop}%
\bibitem [{\citenamefont {Ivanchenko}\ and\ \citenamefont
  {Soboleva}(1990)}]{Ivanchenko1990}%
  \BibitemOpen
  \bibfield  {author} {\bibinfo {author} {\bibfnamefont {Y.}~\bibnamefont
  {Ivanchenko}}\ and\ \bibinfo {author} {\bibfnamefont {T.}~\bibnamefont
  {Soboleva}},\ }\bibfield  {title} {\bibinfo {title} {{Nonlocal interaction in
  Josephson junctions}},\ }\href
  {https://doi.org/https://doi.org/10.1016/0375-9601(90)90015-G} {\bibfield
  {journal} {\bibinfo  {journal} {Phys. Lett. A}\ }\textbf {\bibinfo {volume}
  {147}},\ \bibinfo {pages} {65} (\bibinfo {year} {1990})}\BibitemShut
  {NoStop}%
\bibitem [{\citenamefont {Abdumalikov}\ \emph {et~al.}(2009)\citenamefont
  {Abdumalikov}, \citenamefont {Alfimov},\ and\ \citenamefont
  {Malishevskii}}]{Abdumalikov2009}%
  \BibitemOpen
  \bibfield  {author} {\bibinfo {author} {\bibfnamefont {A.~A.}\ \bibnamefont
  {Abdumalikov}, \bibfnamefont {Jr.}}, \bibinfo {author} {\bibfnamefont
  {G.~L.}\ \bibnamefont {Alfimov}},\ and\ \bibinfo {author} {\bibfnamefont
  {A.~S.}\ \bibnamefont {Malishevskii}},\ }\bibfield  {title} {\bibinfo {title}
  {{Nonlocal electrodynamics of Josephson vortices in superconducting
  circuits}},\ }\href {https://doi.org/10.1088/0953-2048/22/2/023001}
  {\bibfield  {journal} {\bibinfo  {journal} {Supercond. Sci. Technol.}\
  }\textbf {\bibinfo {volume} {22}},\ \bibinfo {pages} {023001} (\bibinfo
  {year} {2009})}\BibitemShut {NoStop}%
\bibitem [{\citenamefont {Boris}\ \emph {et~al.}(2013)\citenamefont {Boris},
  \citenamefont {Rydh}, \citenamefont {Golod}, \citenamefont {Motzkau},
  \citenamefont {Klushin},\ and\ \citenamefont {Krasnov}}]{Boris2013}%
  \BibitemOpen
  \bibfield  {author} {\bibinfo {author} {\bibfnamefont {A.~A.}\ \bibnamefont
  {Boris}}, \bibinfo {author} {\bibfnamefont {A.}~\bibnamefont {Rydh}},
  \bibinfo {author} {\bibfnamefont {T.}~\bibnamefont {Golod}}, \bibinfo
  {author} {\bibfnamefont {H.}~\bibnamefont {Motzkau}}, \bibinfo {author}
  {\bibfnamefont {A.~M.}\ \bibnamefont {Klushin}},\ and\ \bibinfo {author}
  {\bibfnamefont {V.~M.}\ \bibnamefont {Krasnov}},\ }\bibfield  {title}
  {\bibinfo {title} {{Evidence for nonlocal electrodynamics in planar Josephson
  junctions}},\ }\href {https://doi.org/10.1103/PhysRevLett.111.117002}
  {\bibfield  {journal} {\bibinfo  {journal} {Phys. Rev. Lett.}\ }\textbf
  {\bibinfo {volume} {111}},\ \bibinfo {pages} {117002} (\bibinfo {year}
  {2013})}\BibitemShut {NoStop}%
\bibitem [{\citenamefont {Kogan}\ \emph {et~al.}(2001)\citenamefont {Kogan},
  \citenamefont {Dobrovitski}, \citenamefont {Clem}, \citenamefont {Mawatari},\
  and\ \citenamefont {Mints}}]{Kogan2001}%
  \BibitemOpen
  \bibfield  {author} {\bibinfo {author} {\bibfnamefont {V.~G.}\ \bibnamefont
  {Kogan}}, \bibinfo {author} {\bibfnamefont {V.~V.}\ \bibnamefont
  {Dobrovitski}}, \bibinfo {author} {\bibfnamefont {J.~R.}\ \bibnamefont
  {Clem}}, \bibinfo {author} {\bibfnamefont {Y.}~\bibnamefont {Mawatari}},\
  and\ \bibinfo {author} {\bibfnamefont {R.~G.}\ \bibnamefont {Mints}},\
  }\bibfield  {title} {\bibinfo {title} {{Josephson junction in a thin film}},\
  }\href {https://doi.org/10.1103/PhysRevB.63.144501} {\bibfield  {journal}
  {\bibinfo  {journal} {Physical Review B}\ }\textbf {\bibinfo {volume} {63}},\
  \bibinfo {pages} {144501} (\bibinfo {year} {2001})}\BibitemShut {NoStop}%
\bibitem [{\citenamefont {Moshe}\ \emph {et~al.}(2008)\citenamefont {Moshe},
  \citenamefont {Kogan},\ and\ \citenamefont {Mints}}]{Moshe2008}%
  \BibitemOpen
  \bibfield  {author} {\bibinfo {author} {\bibfnamefont {M.}~\bibnamefont
  {Moshe}}, \bibinfo {author} {\bibfnamefont {V.~G.}\ \bibnamefont {Kogan}},\
  and\ \bibinfo {author} {\bibfnamefont {R.~G.}\ \bibnamefont {Mints}},\
  }\bibfield  {title} {\bibinfo {title} {{Edge-type Josephson junctions in
  narrow thin-film strips}},\ }\href
  {https://doi.org/10.1103/PhysRevB.78.020510} {\bibfield  {journal} {\bibinfo
  {journal} {Phys. Rev. B}\ }\textbf {\bibinfo {volume} {78}},\ \bibinfo
  {pages} {020510(R)} (\bibinfo {year} {2008})}\BibitemShut {NoStop}%
\bibitem [{\citenamefont {Clem}(2010)}]{Clem2010}%
  \BibitemOpen
  \bibfield  {author} {\bibinfo {author} {\bibfnamefont {J.~R.}\ \bibnamefont
  {Clem}},\ }\bibfield  {title} {\bibinfo {title} {Josephson junctions in thin
  and narrow rectangular superconducting strips},\ }\href
  {https://doi.org/10.1103/PhysRevB.81.144515} {\bibfield  {journal} {\bibinfo
  {journal} {Phys. Rev. B}\ }\textbf {\bibinfo {volume} {81}},\ \bibinfo
  {pages} {144515} (\bibinfo {year} {2010})}\BibitemShut {NoStop}%
\bibitem [{\citenamefont {Rodan-Legrain}\ \emph {et~al.}(2021)\citenamefont
  {Rodan-Legrain}, \citenamefont {Cao}, \citenamefont {Park}, \citenamefont
  {de~la Barrera}, \citenamefont {Mallika}, \citenamefont {Watanabe},
  \citenamefont {Tanigushi},\ and\ \citenamefont
  {Jarillo-Herrero}}]{Rodan-Legrain2021}%
  \BibitemOpen
  \bibfield  {author} {\bibinfo {author} {\bibfnamefont {D.}~\bibnamefont
  {Rodan-Legrain}}, \bibinfo {author} {\bibfnamefont {Y.}~\bibnamefont {Cao}},
  \bibinfo {author} {\bibfnamefont {J.}~\bibnamefont {Park}}, \bibinfo {author}
  {\bibfnamefont {S.}~\bibnamefont {de~la Barrera}}, \bibinfo {author}
  {\bibfnamefont {T.}~\bibnamefont {Mallika}}, \bibinfo {author} {\bibfnamefont
  {K.}~\bibnamefont {Watanabe}}, \bibinfo {author} {\bibfnamefont
  {T.}~\bibnamefont {Tanigushi}},\ and\ \bibinfo {author} {\bibfnamefont
  {P.}~\bibnamefont {Jarillo-Herrero}},\ }\bibfield  {title} {\bibinfo {title}
  {{Highly tunable junctions and non-local Josephson effect in magic-angle
  graphene tunnelling devices}},\ }\href
  {http://dx.doi.org/10.1038/s41565-021-00894-4
  http://www.nature.com/articles/s41565-021-00894-4} {\bibfield  {journal}
  {\bibinfo  {journal} {Nat. Nanotechnol.}\ }\textbf {\bibinfo {volume} {16}},\
  \bibinfo {pages} {769} (\bibinfo {year} {2021})}\BibitemShut {NoStop}%
\bibitem [{Note1()}]{Note1}%
  \BibitemOpen
  \bibinfo {note} {Here we assume a weak Josephson current, such that the
  magnitude of the superconducting order parameter is not suppressed, and is
  given by the equilibrium value.}\BibitemShut {Stop}%
\bibitem [{com()}]{comment_paper}%
  \BibitemOpen
  \href@noop {} {}\bibinfo {note} {Figure 5a and b are used as a control
  experiment in Ref. 14; the data is presented in the Supporting Information of
  that paper.}\BibitemShut {Stop}%
\bibitem [{\citenamefont {Fermin}\ \emph
  {et~al.}(2022{\natexlab{b}})\citenamefont {Fermin}, \citenamefont
  {Scheinowitz}, \citenamefont {Aarts},\ and\ \citenamefont
  {Lahabi}}]{ellipse_paper2022}%
  \BibitemOpen
  \bibfield  {author} {\bibinfo {author} {\bibfnamefont {R.}~\bibnamefont
  {Fermin}}, \bibinfo {author} {\bibfnamefont {N.~M.~A.}\ \bibnamefont
  {Scheinowitz}}, \bibinfo {author} {\bibfnamefont {J.}~\bibnamefont {Aarts}},\
  and\ \bibinfo {author} {\bibfnamefont {K.}~\bibnamefont {Lahabi}},\
  }\bibfield  {title} {\bibinfo {title} {Mesoscopic superconducting memory
  based on bistable magnetic textures},\ }\href
  {https://doi.org/10.1103/PhysRevResearch.4.033136} {\bibfield  {journal}
  {\bibinfo  {journal} {Phys. Rev. Research}\ }\textbf {\bibinfo {volume}
  {4}},\ \bibinfo {pages} {033136} (\bibinfo {year}
  {2022}{\natexlab{b}})}\BibitemShut {NoStop}%
\bibitem [{\citenamefont {Mandal}\ \emph {et~al.}(2020)\citenamefont {Mandal},
  \citenamefont {Dutta}, \citenamefont {Basistha}, \citenamefont {Roy},
  \citenamefont {Jesudasan}, \citenamefont {Bagwe}, \citenamefont {Benfatto},
  \citenamefont {Thamizhavel},\ and\ \citenamefont
  {Raychaudhuri}}]{Mandal2020}%
  \BibitemOpen
  \bibfield  {author} {\bibinfo {author} {\bibfnamefont {S.}~\bibnamefont
  {Mandal}}, \bibinfo {author} {\bibfnamefont {S.}~\bibnamefont {Dutta}},
  \bibinfo {author} {\bibfnamefont {S.}~\bibnamefont {Basistha}}, \bibinfo
  {author} {\bibfnamefont {I.}~\bibnamefont {Roy}}, \bibinfo {author}
  {\bibfnamefont {J.}~\bibnamefont {Jesudasan}}, \bibinfo {author}
  {\bibfnamefont {V.}~\bibnamefont {Bagwe}}, \bibinfo {author} {\bibfnamefont
  {L.}~\bibnamefont {Benfatto}}, \bibinfo {author} {\bibfnamefont
  {A.}~\bibnamefont {Thamizhavel}},\ and\ \bibinfo {author} {\bibfnamefont
  {P.}~\bibnamefont {Raychaudhuri}},\ }\bibfield  {title} {\bibinfo {title}
  {Destruction of superconductivity through phase fluctuations in ultrathin
  $a$-moge films},\ }\href {https://doi.org/10.1103/PhysRevB.102.060501}
  {\bibfield  {journal} {\bibinfo  {journal} {Phys. Rev. B}\ }\textbf {\bibinfo
  {volume} {102}},\ \bibinfo {pages} {060501(R)} (\bibinfo {year}
  {2020})}\BibitemShut {NoStop}%
\bibitem [{Note2()}]{Note2}%
  \BibitemOpen
  \bibinfo {note} {\label {note1}See the Supplemental Material for a
  description of the technical details of the Fourier transform and, results
  obtained on the bar-shaped sample.}\BibitemShut {Stop}%
\bibitem [{Note3()}]{Note3}%
  \BibitemOpen
  \bibinfo {note} {We assume a relatively high uncertainty in the junction
  length due to the superconducting electrodes shielding the line of sight to
  the junction, which makes establishing $W_{\protect \text {JJ}}$ more
  difficult. Furthermore, we have used $\Delta B = \Delta B_4$ since a rapid
  decay of the $I_{\protect \text {c}}(B)$ pattern prevented establishing
  $\Delta B_5$. It must be noted that the average over the periodicity of the
  side peaks yields $\Delta B W_{\protect \text {JJ}}^2/\Phi _0 =
  1.83$.}\BibitemShut {Stop}%
\bibitem [{\citenamefont {Huang}\ \emph {et~al.}(2019)\citenamefont {Huang},
  \citenamefont {Zhou}, \citenamefont {Zhang}, \citenamefont {Yang},
  \citenamefont {Liu}, \citenamefont {Wang}, \citenamefont {Wan}, \citenamefont
  {Huang}, \citenamefont {Liao}, \citenamefont {Zhang}, \citenamefont {Liu},
  \citenamefont {Deng}, \citenamefont {Chen}, \citenamefont {Han},
  \citenamefont {Zou}, \citenamefont {Lin}, \citenamefont {Han}, \citenamefont
  {Wang}, \citenamefont {Law},\ and\ \citenamefont {Xiu}}]{Huang2019}%
  \BibitemOpen
  \bibfield  {author} {\bibinfo {author} {\bibfnamefont {C.}~\bibnamefont
  {Huang}}, \bibinfo {author} {\bibfnamefont {B.~T.}\ \bibnamefont {Zhou}},
  \bibinfo {author} {\bibfnamefont {H.}~\bibnamefont {Zhang}}, \bibinfo
  {author} {\bibfnamefont {B.}~\bibnamefont {Yang}}, \bibinfo {author}
  {\bibfnamefont {R.}~\bibnamefont {Liu}}, \bibinfo {author} {\bibfnamefont
  {H.}~\bibnamefont {Wang}}, \bibinfo {author} {\bibfnamefont {Y.}~\bibnamefont
  {Wan}}, \bibinfo {author} {\bibfnamefont {K.}~\bibnamefont {Huang}}, \bibinfo
  {author} {\bibfnamefont {Z.}~\bibnamefont {Liao}}, \bibinfo {author}
  {\bibfnamefont {E.}~\bibnamefont {Zhang}}, \bibinfo {author} {\bibfnamefont
  {S.}~\bibnamefont {Liu}}, \bibinfo {author} {\bibfnamefont {Q.}~\bibnamefont
  {Deng}}, \bibinfo {author} {\bibfnamefont {Y.}~\bibnamefont {Chen}}, \bibinfo
  {author} {\bibfnamefont {X.}~\bibnamefont {Han}}, \bibinfo {author}
  {\bibfnamefont {J.}~\bibnamefont {Zou}}, \bibinfo {author} {\bibfnamefont
  {X.}~\bibnamefont {Lin}}, \bibinfo {author} {\bibfnamefont {Z.}~\bibnamefont
  {Han}}, \bibinfo {author} {\bibfnamefont {Y.}~\bibnamefont {Wang}}, \bibinfo
  {author} {\bibfnamefont {K.~T.}\ \bibnamefont {Law}},\ and\ \bibinfo {author}
  {\bibfnamefont {F.}~\bibnamefont {Xiu}},\ }\bibfield  {title} {\bibinfo
  {title} {{Proximity-induced surface superconductivity in Dirac semimetal
  Cd$_3$As$_2$}},\ }\href {https://doi.org/10.1038/s41467-019-10233-w}
  {\bibfield  {journal} {\bibinfo  {journal} {Nat. Commun.}\ }\textbf {\bibinfo
  {volume} {10}},\ \bibinfo {pages} {2217} (\bibinfo {year}
  {2019})}\BibitemShut {NoStop}%
\bibitem [{\citenamefont {Suominen}\ \emph {et~al.}(2017)\citenamefont
  {Suominen}, \citenamefont {Danon}, \citenamefont {Kjaergaard}, \citenamefont
  {Flensberg}, \citenamefont {Shabani}, \citenamefont {Palmstr{\o}m},
  \citenamefont {Nichele},\ and\ \citenamefont {Marcus}}]{Suominen2017}%
  \BibitemOpen
  \bibfield  {author} {\bibinfo {author} {\bibfnamefont {H.~J.}\ \bibnamefont
  {Suominen}}, \bibinfo {author} {\bibfnamefont {J.}~\bibnamefont {Danon}},
  \bibinfo {author} {\bibfnamefont {M.}~\bibnamefont {Kjaergaard}}, \bibinfo
  {author} {\bibfnamefont {K.}~\bibnamefont {Flensberg}}, \bibinfo {author}
  {\bibfnamefont {J.}~\bibnamefont {Shabani}}, \bibinfo {author} {\bibfnamefont
  {C.~J.}\ \bibnamefont {Palmstr{\o}m}}, \bibinfo {author} {\bibfnamefont
  {F.}~\bibnamefont {Nichele}},\ and\ \bibinfo {author} {\bibfnamefont {C.~M.}\
  \bibnamefont {Marcus}},\ }\bibfield  {title} {\bibinfo {title} {{Anomalous
  Fraunhofer interference in epitaxial superconductor-semiconductor Josephson
  junctions}},\ }\href {https://doi.org/10.1103/PhysRevB.95.035307} {\bibfield
  {journal} {\bibinfo  {journal} {Physical Review B}\ }\textbf {\bibinfo
  {volume} {95}},\ \bibinfo {pages} {035307} (\bibinfo {year}
  {2017})}\BibitemShut {NoStop}%
\bibitem [{\citenamefont {de~Vries}\ \emph {et~al.}(2018)\citenamefont
  {de~Vries}, \citenamefont {Timmerman}, \citenamefont {Ostroukh},
  \citenamefont {van Veen}, \citenamefont {Beukman}, \citenamefont {Qu},
  \citenamefont {Wimmer}, \citenamefont {Nguyen}, \citenamefont {Kiselev},
  \citenamefont {Yi}, \citenamefont {Sokolich}, \citenamefont {Manfra},
  \citenamefont {Marcus},\ and\ \citenamefont {Kouwenhoven}}]{Vries2018}%
  \BibitemOpen
  \bibfield  {author} {\bibinfo {author} {\bibfnamefont {F.~K.}\ \bibnamefont
  {de~Vries}}, \bibinfo {author} {\bibfnamefont {T.}~\bibnamefont {Timmerman}},
  \bibinfo {author} {\bibfnamefont {V.~P.}\ \bibnamefont {Ostroukh}}, \bibinfo
  {author} {\bibfnamefont {J.}~\bibnamefont {van Veen}}, \bibinfo {author}
  {\bibfnamefont {A.~J.~A.}\ \bibnamefont {Beukman}}, \bibinfo {author}
  {\bibfnamefont {F.}~\bibnamefont {Qu}}, \bibinfo {author} {\bibfnamefont
  {M.}~\bibnamefont {Wimmer}}, \bibinfo {author} {\bibfnamefont {B.-M.}\
  \bibnamefont {Nguyen}}, \bibinfo {author} {\bibfnamefont {A.~A.}\
  \bibnamefont {Kiselev}}, \bibinfo {author} {\bibfnamefont {W.}~\bibnamefont
  {Yi}}, \bibinfo {author} {\bibfnamefont {M.}~\bibnamefont {Sokolich}},
  \bibinfo {author} {\bibfnamefont {M.~J.}\ \bibnamefont {Manfra}}, \bibinfo
  {author} {\bibfnamefont {C.~M.}\ \bibnamefont {Marcus}},\ and\ \bibinfo
  {author} {\bibfnamefont {L.~P.}\ \bibnamefont {Kouwenhoven}},\ }\bibfield
  {title} {\bibinfo {title} {{$h/e$ superconducting quantum interference
  through trivial edge states in InAs}},\ }\href
  {https://doi.org/10.1103/PhysRevLett.120.047702} {\bibfield  {journal}
  {\bibinfo  {journal} {Phys. Rev. Lett.}\ }\textbf {\bibinfo {volume} {120}},\
  \bibinfo {pages} {047702} (\bibinfo {year} {2018})}\BibitemShut {NoStop}%
\bibitem [{Note4()}]{Note4}%
  \BibitemOpen
  \bibinfo {note} {Naturally, any choice of $\protect \tilde {y}$ and $\protect
  \tilde {\beta }$ is allowed, as long as it is consistent with $\gamma
  $.}\BibitemShut {Stop}%
\end{thebibliography}

\end{document}


\title{Supplemental Material of: Beyond the effective length: How to analyze magnetic interference patterns of thin film planar Josephson junctions with finite lateral dimensions}

\date{\today}

\author{R. Fermin}
\author{B. de Wit}
\author{J. Aarts}
\affiliation{Huygens-Kamerlingh Onnes Laboratory, Leiden University, P.O. Box 9504, 2300 RA Leiden, The Netherlands.}

\pacs{} \maketitle

\clearpage
 
\section{Data obtained on a bar-shaped sample}\label{Bar_sample}

\begin{figure*}[h!]
 \centerline{$
 \begin{array}{c}
  \includegraphics[width=1\linewidth]{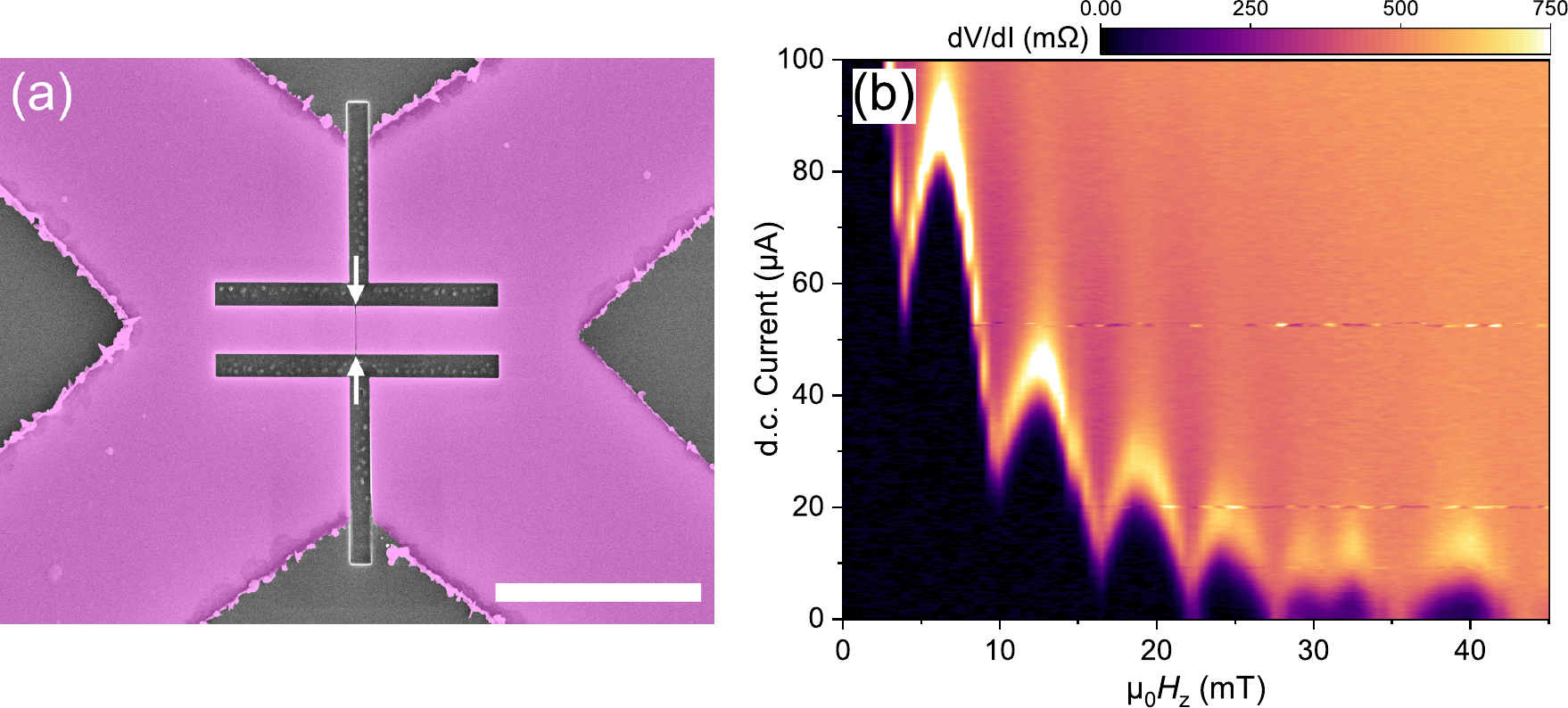}
 \end{array}$}
 \caption{(a) Scanning electron micrograph of a bar-shaped sample for which $L/W > 10$. The arrows indicate the location of the junction and the scale bar corresponds to 5 {\textmu}m. (b) $dV/dI$ color map of the magnetic interference pattern ($I_{\text{c}}(B)$) obtained on the junction shown in (a). Note that the oscillation amplitude decreases relatively rapidly with respect to the ellipse-shaped samples. Therefore it becomes impossible to establish the fifth side lobe ($\Delta B_5$) of the pattern.} \label{Bar_sample_fig}
\end{figure*}

In the main text we present results obtained on elliptical samples, in this Supplemental Material we show the results obtained on a rectangular junction. In Figure \ref{Bar_sample_fig} a scanning electron micrograph of this junctions accompanies a color plot of the $I_{\text{c}}(B)$-pattern obtained on this sample. We have used this sample to examine the geometrical independent limit where $L \gg W$. Since the oscillation amplitude of the bar-shaped sample decreases more strongly with increasing $B$ than the ellipse-shaped samples, we cannot determine $\Delta B$ using the fifth side lobe of the pattern. Instead, we use the width of the fourth side lobe to establish $\Delta B$. Following the discussion in the main text this yields: $\Delta B W_{\text{JJ}}^2/\Phi_0 = 1.70$. Interestingly, if we calculate $\Delta B$ by taking the average of the first four side lobes in the interference pattern, we find $\Delta B W_{\text{JJ}}^2/\Phi_0 = 1.83$.
 
\clearpage

\section{Technical details of the Fourier analysis}\label{tech_detail_Fourier}

This Supplemental Material will give the details of carrying out the Fourier transform described in the main text. It concerns an updates version of the material published as Supporting Information of reference 14 of the main text. 

The Fourier transform is defined in the main text as:

\begin{align} \label{Eq:H3_8_sub}
I_{\text{c}}(\Tilde{\beta}) =  \left| \int_{-\infty}^{\infty} \Tilde{J}_{\text{c}}(\Tilde{y}) e^{i2\pi \Tilde{\beta} \Tilde{y}} \: \text{d}\Tilde{y} \right| = \left| \mathfrak{I}_{\text{c}} \right|
\end{align}

In the right hand side, the transform $\mathfrak{I}_{\text{c}}$ is complex valued; its real and imaginary parts encode for the even and odd components in $I_c(\Tilde{\beta})$ respectively. Since the experimental interference patterns are mainly symmetric (i.e., an even function of the applied magnetic field), we can assume $\mathfrak{I}_{\text{c}}$ to be dominantly real:

\begin{align} \label{Eq:H3_11}
I_{\text{c},\text{even}}(\Tilde{\beta}) =   \int_{-\infty}^{\infty} \Tilde{J}_{\text{c},\text{even}}(\Tilde{y}) \cos(\Tilde{\beta} \Tilde{y})  \: \text{d}\Tilde{y}
\end{align}

\noindent The real component of $\mathfrak{I}_{\text{c}}$ is an oscillating function that flips sign at each zero crossing. The imaginary part is significantly smaller than the real part, except at the zero-crossing where the even part vanishes. Therefore, the imaginary part of $\mathfrak{I}_{\text{c}}$ ($I_{\text{c},\text{odd}}(\Tilde{\beta})$) can be approximated by the critical current at the minima in the experimental interference pattern. Also $I_{\text{c},\text{odd}}(\Tilde{\beta})$ is flipping its sign between each minimum and between the minima we approximate $I_{\text{c},\text{odd}}(\Tilde{\beta})$ by linear interpolation. The inverse transform yielding $\Tilde{J}_{\text{c}}(\Tilde{y})$ from $I_{\text{c}}(\Tilde{\beta})$ is then given by:

\begin{align} \label{Eq:H3_12}
\Tilde{J}_{\text{c}}(\Tilde{y}) =  \left|  \int_{-\infty}^{\infty} \left( I_{\text{c},\text{even}}(\Tilde{\beta}) + i I_{\text{c},\text{odd}}(\Tilde{\beta}) \right) e^{-i\Tilde{\beta} \Tilde{y}} \: \text{d}\Tilde{\beta} \right| = \left|  \int_{-\infty}^{\infty} \mathfrak{I}_{\text{c}} \: e^{-i\Tilde{\beta} \Tilde{y}} \: \text{d}\Tilde{\beta} \right|
\end{align}

\begin{figure*}[t!]
 \centerline{$
 \begin{array}{c}
  \includegraphics[width=1\linewidth]{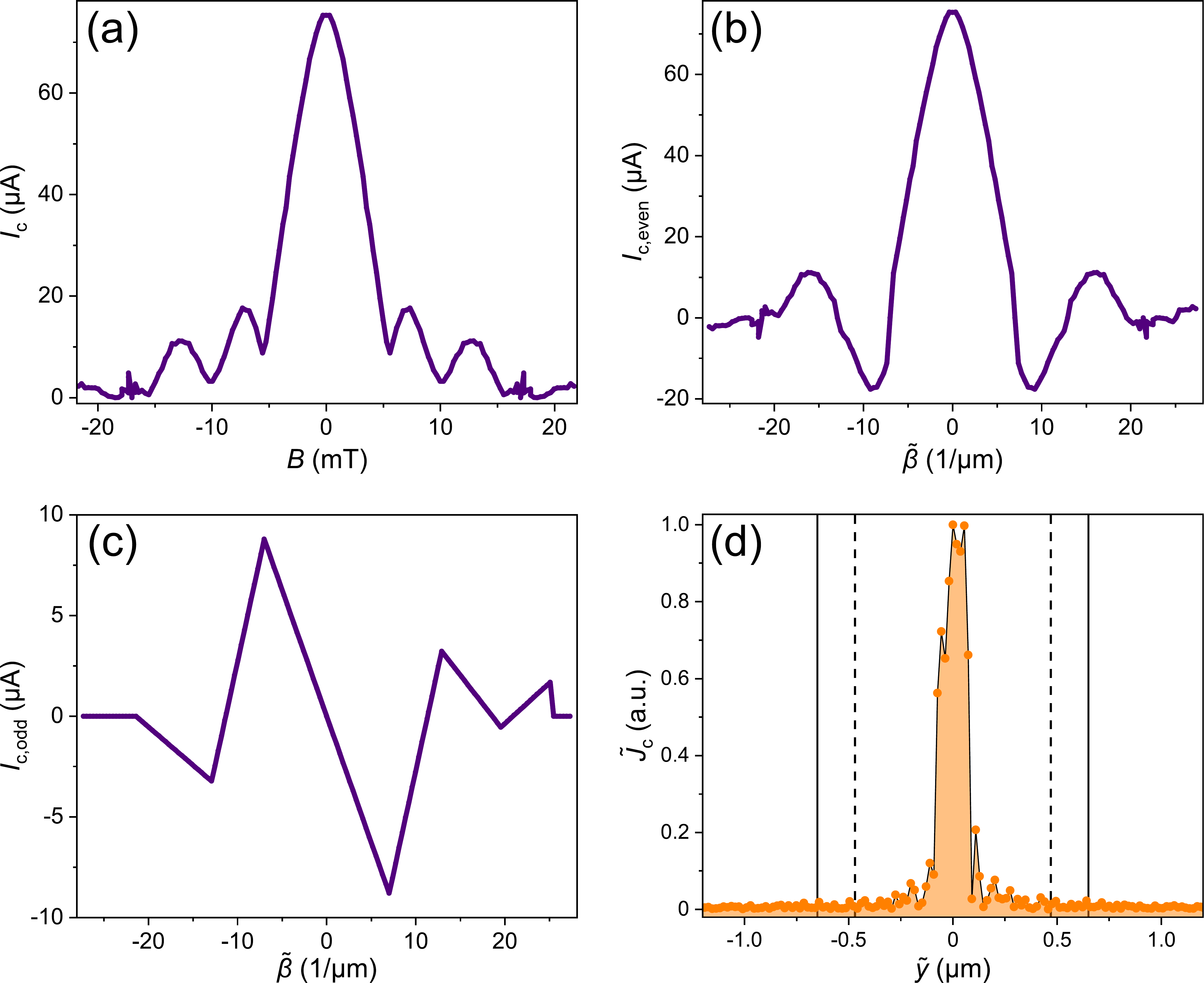}
 \end{array}$}
 \caption{Overview of the subsequent steps of the Fourier transform analysis. (a) Extracted critical current on the basis of a voltage cut-off. In order to not overestimate the imaginary part of the complex critical current, the global minimum of the interference pattern is shifted to zero current. (b) and (c) respectively show the real and imaginary part of the complex critical current extracted from the interference pattern depicted in (a). Here the subsequent lobes are flipped to obtain the real part, and the minima are interpolated to obtain the imaginary part. (d) The resulting critical current density distribution, calculated using the Fourier transform described in Eq. \ref{Eq:H3_12}. The result of (d) is scaled using $f_{\text{disk}}\left(\frac{y}{W}\right)$, to retrieve the actual $J_{\text{c}}(y)$, which shown in the Figure 5a of the main text. The boundaries of the electrodes ($-W/2$ and $W/2$) are indicated by solid reference lines and the boundaries of the actual weak link ($-W_{\text{JJ}}/2$ and $W_{\text{JJ}}/2$) by dotted reference lines.} \label{H3_Fourier}
\end{figure*}

Figure \ref{H3_Fourier} gives an overview of the subsequent steps of the Fourier analysis. First, $I_{\text{c}}$ is extracted from the experimental data by a defining a voltage threshold; this is depicted in Figure \ref{H3_Fourier}a for a disk-shaped junction (data in Figure 5b of the main text). We vertically translate the extracted $I_{\text{c}}$ values such that the global minimum equals zero current. This step in the data analysis prevents the overestimation of $I_{\text{c},\text{odd}}(\Tilde{\beta})$, which would result in an overly anti-symmetric current density distribution. $I_{\text{c},\text{even}}(\Tilde{\beta})$ is found by multiplying the translated $I_{\text{c}}$ by a flipping function that changes the sign of each subsequent lobe of the interference pattern, as can be observed in Figure \ref{H3_Fourier}b.

\clearpage

Here we also rescale the field axis to $\Tilde{\beta}$. We follow the above procedure for finding $I_{\text{c},\text{odd}}(\Tilde{\beta})$, which is depicted in Figure \ref{H3_Fourier}c. The corresponding critical current density distribution is found by a numerical Fourier transform carried out in Python using the Numpy package, yielding the distribution $\Tilde{J}_{\text{c}}(\Tilde{y})$, depicted in Figure \ref{H3_Fourier}d. Finally both axes are rescaled using $f_{\text{disk}}\left(\frac{y}{W}\right)$ to retrieve $J_{\text{c}}(y)$, which is shown in Figure 7a of the main text. For illustrative purposes, we have chosen to absorb part of the prefactor in $\Tilde{y}$, as this this yields a larger contrast between Figure \ref{H3_Fourier}d and Figure 7a of the main text. However, as discussed in the main text, any choice of $\Tilde{y}$ and $\Tilde{\beta}$ is allowed, as long as it is consistent with $\gamma$. We indicate $-W/2$ and $W/2$ by solid reference lines and the boundaries of the actual weak link by dotted reference lines. For the Fourier transform using a linear approximation of $f_{\text{disk}}\left(\frac{y}{W}\right)$ we define $\Tilde{y} = y$ and absorb the fit of $f_{\text{disk}}\left(\frac{y}{W}\right)$ into $\Tilde{\beta}$.

%
